\documentclass[11pt]{article}
\usepackage{amssymb}
\usepackage{graphics}
\usepackage{epsfig}
\usepackage{a4wide}
\usepackage{color}

\begin{document}

\newcommand{\ttbs}{\char'134}
\newcommand\fverb{\setbox\fverbbox=\hbox\bgroup\verb}
\newcommand\fverbdo{\egroup\medskip\noindent%
            \fbox{\unhbox\fverbbox}\ }
\newcommand\fverbit{\egroup\item[\fbox{\unhbox\fverbbox}]}
\newbox\fverbbox
\newcommand{\nb}{\nonumber}
\newcommand{\ba}{\begin{array}}
\newcommand{\ea}{\end{array}}
\newcommand{\gdtsl}{$gd_k \to t\slep_i^-$ }
\newcommand{\pptsl}{$pp \to t\slep_i+X~$}
\newcommand{\realgluon}{$gd_k \to t\slep_i+g$}
\newcommand{\tabincell}[2]{\begin{tabular}{@{}#1@{}}#2\end{tabular}}

\def\rp{R\!\!\!/ _p}
\def\slep{\tilde{\ell}}

\title{ Single (anti-)top quark production in association with a lightest neutralino
at LHeC }

\author{ Li Xiao-Peng$^{a}$, Guo Lei$^{a}$, Ma Wen-Gan$^{a}$,
Zhang Ren-You$^{a}$, Han Liang$^{a}$, and Song Mao$^{b}$  \\
{\small $^a$ Department of Modern Physics, University of Science and
Technology of China,} \\
{\small  Hefei, Anhui 230026, People's Republic of China} \\
{\small $^b$ School of Physics and Material Science, Anhui University,
Hefei, Anhui 230039,  People's Republic of China  }}

\date{}
\maketitle \vskip 15mm
\begin{abstract}
In this paper we examine thoroughly the single (anti-)top quark
production associated with a lightest neutralino at the possible
CERN Large Hadron Electron Collider (LHeC). We calculate the full
next-to-leading order (NLO) QCD corrections to the process including all
the nonresonant diagrams and do not assume the production and decay
factorization for the possible resonant top squarks in
the $R$-parity violating minimal supersymmetric standard model
with 19 unrelated parameters. We investigate numerically the
effects of the relevant supersymmetry (SUSY) parameters on the cross
section, and present the transverse momentum distributions of final
(anti-)top quark at both the leading-order (LO) and the QCD NLO. We
find that the NLO QCD corrections enhance the LO cross sections in
most chosen parameter space, and the NLO QCD impacts on the
transverse momentum distributions of the final (anti-)top quark might be
resolvable in some cases. We conclude that even with recently
known experimental constraints the on SUSY parameters, the production
rate could still achieve notable value in the admissible parameter
space.
\end{abstract}

\vskip 25mm

\vskip 15mm {\large\bf PACS: 12.38.Bx, 14.65.Ha, 12.60.Jv}
\vfill \eject

\section{Introduction}
\par
The ATLAS and CMS experiments have collected about $5~fb^{-1}$ data
at the $7~TeV$ LHC and $21 fb^{-1}$ data at the $8~TeV$ LHC. However
the present results searching for supersymmetry (SUSY) have shown no
significant signature \cite{exbounds1,exbounds2}. The SUSY bounds
depend on many details; it is fair to say that the gluino and the
squarks of the first two generations have to be heavier than
$1~TeV$, while the third generation squarks are still can be lighter
than $1~TeV$ in some SUSY models, such as the natural minimal
supersymmetric standard model (nMSSM) and the phenomenological minimal
supersymmetric standard model (pMSSM) \cite{SusyStatus}. In the
nMSSM, gluinos, Higgsinos and the third generation squarks should be
light because of the natural electroweak symmetry breaking request,
and the nMSSM has not been obviously constrained by the LHC
experiment data \cite{naturesusy}. The pMSSM with 19 free parameters
(pMSSM-19) produces a broad perspective on SUSY phenomenology and
gives pMSSM greater flexibility than other models \cite{higgsfit}.
So there is ample room in the pMSSM-19 parameter space which can
survive from the early phase of the LHC experiment \cite{pMSSM1}.

\par
In the minimal supersymmetric standard model (MSSM), the
conservation of $R$-parity is phenomenologically desirable, but is
$ad~hoc$ in the sense that it is not required for the internal
consistency of the theory \cite{Martin}. The generic MSSM without
discrete $R$-parity symmetry \cite{rp1,rp2} conservation has the
advantage of a richer phenomenology, including neutrino masses and
mixing without introducing any extra superfields. The $R$-parity
symmetry implies a conserved quantum number, $R=(-1)^{3B+L+2S}$,
where $B$, $L$ and $S$ are the baryon number, lepton number and spin of
the particle, respectively. $R$ is equal to $+1$ for all the
standard model (SM) particles, while for all the superpartners $R$
is $-1$. Thus the conservation of $R$-parity requests that the
superpartners can only be produced in pairs and the lightest SUSY
particle is stable. However, such a stringent symmetry appears to be
short of a theoretical basis and there is not enough experimental
evidence for $R$-parity conservation. Moreover, the SUSY particle
can be produced singly in $R$-parity violation (RPV) scenarios, and
it would have lower energy threshold than the pair production of the
SUSY particle in the $R$-parity conserving MSSM. In the following we
consider the pMSSM extended to include RPV interactions. The most
general superpotential with RPV interactions can be written as
\cite{superpotential_1,superpotential_2,superpotential_3}
\begin{equation}\label{Lagrangian-1}
{\cal W}_{\rlap/R_{p}} = \frac{1}{2}\epsilon_{ab}
\lambda_{ijk}\hat{L}_{i}^a \hat{L}_{j}^b \hat{E}_{k} +
\epsilon_{ab}\lambda^{'}_{ijk} \hat{L}_{i}^a \hat{Q}_{j}^b
\hat{D}_{k} +
\frac{1}{2}\epsilon_{\alpha\beta\gamma}\lambda^{''}_{ijk}
   \hat{U}_{i}^{\alpha} \hat{D}_{j}^{\beta} \hat{D}_{k}^{\gamma} +
\epsilon_{ab}\mu_{i} \hat{L}_{i}^a \hat{H}_{2}^b,
\end{equation}
where $i$, $j$, $k$ denote generation indices, $a,b~(= 1,2)$ are
$SU(2)$ isospin indices, and $\alpha$, $\beta$, $\gamma$ are $SU(3)$
color indices. $\hat{L}$ ($\hat{Q}$) are the left-handed lepton
(quark) $SU(2)$ doublet chiral superfields, and $\hat{E}$
($\hat{U}$, $\hat{D}$) are the right-handed lepton (up- and
down-type quark) $SU(2)$ singlet chiral superfields. $H_{2}$ is one
of the Higgs chiral superfields. $\lambda_{ijk}$ and
$\lambda^{\prime}_{ijk}$ are the $L$-violating dimensionless
coupling coefficients, and $\lambda^{\prime\prime}_{ijk}$ are the
$B$-violating dimensionless coupling constants. We know that a
stable proton can survive by imposing that $B$ and $L$ cannot be
violated at the same time \cite{LBviolation}.

\par
In this work we focus on the single (anti-)top quark production
associated with a lightest neutralino at the proposed Large Hadron
Electron Collider (LHeC), which provides a complement to the LHC by
using the existing $7~TeV$ proton beam and $E_e = 50 \sim 150~GeV$
electron (positron) beam \cite{LHeC}. If the $L$-violating coupling
coefficients, $\lambda^{\prime}_{13k}$, are nonzero, the possible
direct resonant scalar top (stop) quark production could enhance the
production rate for the single top quark production process.

\par
The $t\tilde{\chi}_1^0$ ($\bar t\tilde{\chi}_1^0$) production process
in the $R$-violating MSSM at the LHeC can be induced by the resonant top squark production
and its subsequent decay, due to the RPV couplings between electron, top squark
and quark, which are from the nonzero term
$\lambda^\prime_{13k} \hat{L}_{1} \hat{Q}_{3} \hat{D}_{k}$
in Eq.(\ref{Lagrangian-1}) and written explicitly as
\begin{equation}\label{Lagrangian-2}
\ba{rcl} {\cal L}_{LQD} & = & \lambda^\prime_{13k} \bigg[
\tilde{\nu}_{1L} \bar{d}_{kR} d_{3L}
+  \tilde{d}_{3L} \bar{d}_{kR} \nu_{1L}
+ \tilde{d}_{kR}^* \overline{\nu_{1L}^c} d_{3L} \bigg. \\
&& \bigg. -  \slep_{1L} \bar{d}_{kR} u_{3L} - \tilde{u}_{3L}
\bar{d}_{kR} \ell_{1L} - \tilde{d}_{kR}^* \overline{\ell_{1L}^c}
u_{3L} \bigg] + ~ \textrm{h.c}. .\ea
\end{equation}

\par
There exist some works on the single top squark production via RPV
coupling at HERA \cite{hera1,hera2,hera3,hera4,hera5} and other
colliders \cite{work1,work2,work3,work4,work5}. In Ref.\cite{hera1},
the $ep\to \tilde t_1$ and the subsequent $\tilde t_1$ decay
processes at the HERA have been studied at the tree level. The
next-to-leading (NLO) QCD correction to the resonant scalar
leptoquark (or squark) production at the HERA was studied in
Ref.\cite{hera5}. The $t\tilde\chi^0_1 (\bar t\tilde\chi^0_1)$
events at the LHeC are produced not only by the single top squark
production and followed by its subsequent decay, but also by
virtual squark and slepton exchanges. Normally, the treatment
considering only the resonant top squark contributions as used in
Ref.\cite{hera5}, is a reasonable approach for the $e^+p(e^-p)\to
t\tilde\chi^0_1 (\bar t\tilde\chi^0_1)+X$ process at the LHeC. But
in some SUSY parameter space with relatively heavy top squark and
relatively light selectron, the contributions from the $t-$ and
$u-$channel diagrams for the process cannot be neglected, particularly
in the NLO QCD precision calculations. In this paper, we focus on
the complete contributions from all diagrams including the single
top squark production mechanism followed with a subsequent decay
process up to NLO at the LHeC in the framework of the $R$-parity
violating pMSSM-19, and do not assume the production and decay
factorization for the possible resonant top squarks. The paper is
organized as follows: In Sect. 2, we present the calculations for
the relevant partonic process at both the LO and the QCD NLO. In
Sect. 3, we give some numerical results and discussion. Finally, a
short summary is given.

\vskip 5mm
\section{Calculation framework }
\subsection{ LO calculations }
\par
The $e^-p(e^+p)\to \bar t\tilde\chi^0_1(t\tilde\chi^0_1)+X$ process
at the LHeC receives the contributions from the partonic processes
$e^-\bar d_k(e^+d_k)\to \bar
t\tilde\chi^0_1(t\tilde\chi^0_1)~(k=1,2)$, where $k$ is the
generation index. The $2\sigma$ upper bounds on
$\lambda^{\prime}_{13k}$ originated from Ref.\cite{rpvbounds} are
listed in Table \ref{tab-1}. There the coefficient
$\lambda^{\prime}_{133}<{\cal O}(10^{-4})$ is in response to the
requirement of $m_{\nu}<1~eV$ and left-right mixing in the sbottom
sector \cite{rpvbounds,bounds}. Considering the strong constraint on
$\lambda^{\prime}_{133}$ in Table \ref{tab-1} and the low
(anti-)bottom luminosity in the parton distribution function (PDF) of
the proton, there cannot be significant production via the $e^-\bar
b(e^+b)\to \bar t\tilde\chi^0_1(t\tilde\chi^0_1)$ partonic process,
so we ignore the contribution from the initial (anti-)bottom quark
in the following calculation.
\begin{table}[!hbp]
\begin{center}
\begin{tabular}{|c|c|}
\hline
$\lambda^\prime_{131}$ & $0.03\times(m_{\tilde t_{L}}/100GeV)$ \\
\hline
$\lambda^\prime_{132}$ & $0.28\times (m_{\tilde t_{L}}/100GeV)$ \\
\hline
$\lambda^\prime_{133}$ & $\mathcal{O}(10^{-4})$ \\
\hline
\end{tabular}
\end{center}
\caption{ \label{tab-1} $2\sigma$ upper bounds on
$\lambda^\prime_{13k}$, where $m_{\tilde{t}_{L}}$ is the mass of the
left-handed top squark $\tilde{t}_{L}$. }
\end{table}

\par
\begin{figure*}
\begin{center}
\includegraphics[ scale = 0.75 ]{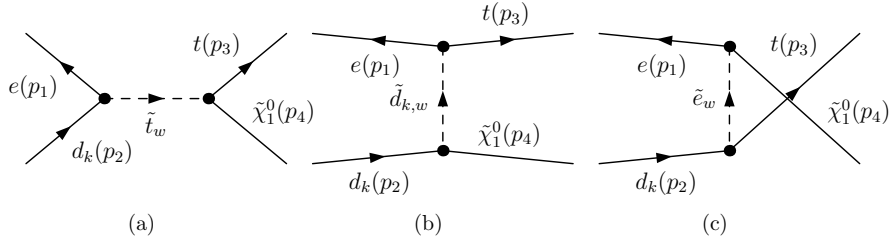}
\caption{\label{fig1} The tree-level Feynman diagrams for the
$e^+d_k\to t\tilde\chi^0_1$ partonic process, where the lower
indices $k,w=1,2$. }
\end{center}
\end{figure*}

In this section we present only the calculations for the $e^+d_k\to
t\tilde\chi^0_1~(k=1,2)$ subprocess which has the same cross section
for $e^-\bar{d}_k\to \bar t\tilde\chi^0_1$ subprocess due to the
$CP$ conservation. The tree-level Feynman diagrams for the
$e^+(p_1)d_k(p_2)\to t(p_3)\tilde\chi^0_1(p_4)$ partonic process are
shown in Fig.\ref{fig1}. The two top squarks
($\tilde{t}_w=\tilde{t}_{1,2}$) in Fig.\ref{fig1}(a) are potentially
resonant. At the LHeC the main contribution to the $e^-p(e^+p)\to
\bar t\tilde\chi^0_1(t\tilde\chi^0_1)+X$ process is from the
$s-$channel diagrams with top squark exchanges, and the contributions
from the $t-$ and $u-$channel diagrams normally are small. But in some
SUSY parameter space where the top squarks are relatively heavy and
the selectrons are relatively light, the u- and t-channel
contributions normally cannot be neglected, particularly in the NLO
QCD precision calculations. For disposal of the singularities due to
stop quark resonances in the calculations, the complex mass scheme
(CMS) is adopted \cite{cms}. In the CMS approach the complex masses
for all related unstable particles should be taken everywhere in
both tree-level and one-loop level calculations. Then the gauge
invariance is kept and the real poles of propagators are avoided. We
introduce the decay widths of $\tilde{t}_{1}$ and $\tilde{t}_{2}$,
and make the following replacements in the amplitudes:
\begin{eqnarray}
\frac{1}{\hat{s}-m_{\tilde{t}_i}^2} \to
\frac{1}{\hat{s}-m_{\tilde{t}_i}^2+i
m_{\tilde{t}_i}\Gamma_{\tilde{t}_i}}=\frac{1}{\hat{s}-\mu_{\tilde{t}_i}^2},
~~(i=1,2), \label{Replace}
\end{eqnarray}
where $\Gamma_{{\tilde{t}_i}}$ represents the decay width of
$\tilde{t}_i$, and $\mu_{\tilde{t}_i}^2$ is the complex mass squared
of $\tilde{t}_i$ defined as $\mu_{\tilde{t}_i}^2=
m_{\tilde{t}_i}^2-i m_{\tilde{t}_i} \Gamma_{\tilde{t}_i}$.

\par
The LO cross section for the partonic process $e^+d_k\to
t\tilde\chi^0_1$ can be written as
\begin{eqnarray}\label{eq-1}
\hat{\sigma}_{0}(e^+d_k\to t\tilde\chi^0_1)&=&
\frac{1}{4}\frac{1}{3}\frac{1}{2\hat{s}}\int \sum_{spin}^{color}
|{\cal M}_{LO}(e^+d_k\to t\tilde\chi^0_1)|^2 d\Omega_{2},
\end{eqnarray}
where the factors $\frac{1}{4}$ and $\frac{1}{3}$ come from the
averaging over the spins and colors of the initial partons
respectively, $\hat{s}$ is the partonic center-of-mass energy
squared, and ${\cal M}_{LO}(e^+d_k\to t\tilde\chi^0_1)$ is the amplitude
for the tree-level Feynman diagrams shown in Fig.\ref{fig1}. The
summation in Eq.(\ref{eq-1}) is taken over the spins and colors of
all the relevant initial and final particles. The phase space
element $d\Omega_2$ is expressed as
\begin{equation}
{d\Omega_{2}}=(2 \pi )^4\delta^{(4)} \left( p_1+p_2-p_3-p_4 \right)
\prod_{i=3,4} \frac{d^3 \textbf{\textsl{p}}_i}{(2 \pi)^3 2 E_i}.
\end{equation}
The LO cross section for the parent process $e^+p\to
t\tilde\chi^0_1+X$ at the LHeC can be obtained by performing the
following integrations:
\begin{equation} \label{LOsigma}
\sigma_{LO} = \sum_{k=1,2}\int_{0}^{1} dx \hat{\sigma}_{0}(e^+d_k\to
t\tilde\chi^0_1) \left[ G_{d_k /P}(x,\mu_f)\right],
\end{equation}
where $G_{d_k /P}(x,\mu_f)$ is the PDF of parton $d_k~(k=1,2)$ in
proton $P$, which describes the probability to find a parton $d_k$
with momentum $xp_P$ in proton $P$, and $\mu_f$ is the factorization
energy scale.

\par
\subsection{ NLO QCD correction }
\par
\begin{figure*}
\begin{center}
\includegraphics[ scale = 0.75 ]{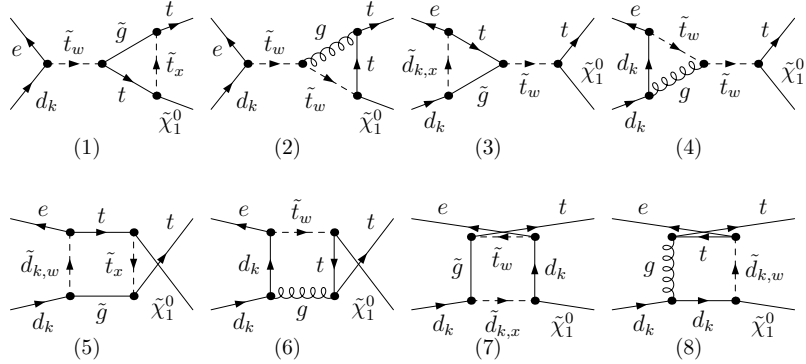}
\caption{\label{fig2} Some representative QCD one-loop Feynman
diagrams for the $e^+d_k\to t\tilde\chi^0_1$ partonic process. The
lower indices $x,w=1,2$ are for distinguishing two squark mass
eigenstates, and the lower index $k$ runs from the first generation
to the second generation. }
\end{center}
\end{figure*}

We use the dimensional regularization scheme in $D=4-2\epsilon$
dimensions to isolate the UV and IR singularities in the NLO
calculations. This scheme violates supersymmetry because the numbers
of the gauge-boson and gaugino degrees of freedom are not equal in
$D\neq 4$ dimensions. To subtract the contributions of the false,
nonsupersymmetric degrees of freedom and restore supersymmetry at
one-loop order, a shift between the $q\tilde{q}\tilde{g}$ Yukawa
coupling $\widehat{g}_s$ and the $qqg$ gauge coupling $g_s$ must be
introduced \cite{shiftgs}:
\begin{equation}\label{gs}
\label{shift} \widehat{g}_s = g_s\left [1 +
\frac{\alpha_s}{8\pi}\left (\frac{4}{3}C_A - C_F\right )\right ],
\end{equation}
where $C_A = N = 3$ and $C_F = 4/3$. In our numerical calculations,
we take this shift into account. Some representative virtual QCD
one-loop diagrams are shown in Fig.\ref{fig2}.

\par
The NLO QCD corrections to the $e^+d_k\to t\tilde\chi^0_1$ partonic
process can be divided into two components: virtual correction and
real radiation correction. There exist UV and 
IR singularities in the virtual correction and only IR singularity
in the real radiation correction. The IR singularity includes soft
IR divergence and collinear IR divergence. In the virtual correction
component, the UV divergence vanishes by performing a renormalization
procedure, and the soft IR divergence can be completely eliminated
by adding the contributions from the real gluon emission partonic
process $e^+d_k\to t\tilde\chi^0_1 g$. The collinear divergence in
the virtual correction can be partially canceled by the collinear
divergence in real emission processes, and the residual collinear
divergence will be absorbed by the redefinitions of the PDFs.

\par
The counterterms of relevant masses and wave functions are defined
as
\begin{eqnarray}\
\label{eq-dectm1} && m_{q,0} = m_q+\delta m_q,~~~~~
m_{\tilde d_i,0} = m_{\tilde d_i}+\delta m_{\tilde d_i},\\
\label{eq-dectm2} && m_{\tilde s_i,0}= m_{\tilde s_i}+\delta
m_{\tilde s_i},~~~~~
\mu^2_{\tilde t_i,0} = \mu^2_{\tilde t_i}+\delta \mu^2_{\tilde t_i},~~(i=1,2)\\
\label{eq-dectf} && q_{L/R,0}=(1+\frac{1}{2}\delta
Z^{q}_{L/R})q_{L/R},~~~~~
\lambda^{\prime}_{13k,0} =\lambda^\prime_{13k}+ \delta\lambda^\prime_{13k},\\
\label{eq-dectsf} && \left(\begin{array}{c} \tilde{q}_{1,0} \\
\tilde{q}_{2,0}\end{array} \right)
=\left(\begin{array}{cc} 1+\frac{1}{2}\delta Z^{\tilde{q}}_{11} & \frac{1}{2}\delta Z^{\tilde{q}}_{12} \\
\frac{1}{2}\delta Z^{\tilde{q}}_{21} & 1+\frac{1}{2}\delta
Z^{\tilde{q}}_{22}
\end{array} \right)\left(\begin{array}{c} \tilde{q}_1 \\
\tilde{q}_2 \end{array} \right),\\
&& \label{eq-dectU} U^{\tilde q}_{0} = U^{\tilde q}+\delta U^{\tilde
q}
=\left(\begin {array}{cc} U^{\tilde q}_{11} & U^{\tilde q}_{12}\\
U^{\tilde q}_{21} & U^{\tilde q}_{22}  \end
{array}\right)+\left(\begin {array}{cc} \delta U^{\tilde q}_{11} &
\delta U^{\tilde q}_{12}\\ \delta U^{\tilde q}_{21} & \delta
U^{\tilde q}_{22} \end {array}\right),
\end{eqnarray}
where $q_{L/R}$, $\tilde q_i$ denote the renormalized fields of
quarks $(d,s,t)$ and squarks $(\tilde d,\tilde s,\tilde t)$ involved
in this process, respectively. $U^{\tilde q}_0$ and $U^{\tilde q}$
represent the bare and renormalized $2\times 2$ mixing matrix of
$\tilde q$. $\mu_{\tilde t_i}$ is the renormalized complex mass of
the top squark $\tilde t_i$. Since we define the top squark masses as
complex ones to deal with the possible top squark resonance at both LO and
NLO, the renormalized mixing matrix elements $U^{\tilde
t}_{ij}~(i,j=1,2)$ might be complex, too.

\par
We apply the CMS in both the LO and the NLO calculations, and
renormalize the relevant fields and corresponding masses in the
on-shell (OS) scheme \cite{os}. For the renormalization of the
$R$-parity violating coupling coefficient $\lambda^{\prime}_{13k}$,
we use the $\overline {MS}$ scheme \cite{msbar}. Due to adopting the
complex top squark masses in our calculation, the one-loop integrals
must be prolongated onto the complex plane continuously. We extend
the formulas for IR-divergent 3,4-point integrals in
Ref.\cite{sol4pi}, and the expressions for IR-safe $N$-point
($N=1,2,3,4$) integrals presented in Ref.\cite{irsafe} analytically
to the complex plane. With the renormalization conditions of the
complex OS scheme \cite{cms,dpf}, the renormalization constants of
the complex masses and wave functions of top squarks are expressed
as
\begin{eqnarray}\label{eq2}
\delta\mu^2_{\tilde{t}_i}  =
\Sigma^{\tilde{t}}_{ii}(\mu_{\tilde{t}_i}^2) ,~~~~ \delta
Z_{ii}^{\tilde{t}} = -
\Sigma_{ii}^{\tilde{t}\prime}(\mu_{\tilde{t}_i}^2),~~~~\delta
Z_{ij}^{\tilde{t}} =
\frac{2\Sigma_{ij}^{\tilde{t}}(\mu_{\tilde{t}_j}^2)}
{\mu_{\tilde{t}_i}^2-\mu_{\tilde{t}_j}^2},~~(~i,j=1,2, i\neq j).
\end{eqnarray}

\par
By performing Taylor series expansion about a real argument for the
top squark self-energy, we have
\begin{eqnarray}\label{eq3}
\Sigma_{ij}^{\tilde{t}}(\mu_{\tilde{t}_j}^2)&=&\Sigma_{ij}^{\tilde{t}}
(m_{\tilde{t}_j}^2)+(\mu_{\tilde{t}_j}^2-m_{\tilde{t}_j}^2)
\Sigma_{ij}^{\tilde{t}\prime}(m_{\tilde{t}_j}^2)\;+\; {\cal O}\left(
(\mu_{\tilde{t}_j}^2-m_{\tilde{t}_j}^2)^2\right)~~\nb \\
&=&\Sigma_{ij}^{\tilde{t}}(m_{\tilde{t}_j}^2)-i
\textit{m}_{\tilde{t}_j}\Gamma_{\tilde{t}_j}
\Sigma_{ij}^{\tilde{t}\prime}(m_{\tilde{t}_j}^2)\;+\; {\cal
O}\left((\textit{m}_{\tilde{t}_j}\Gamma_{\tilde{t}_j})^2\right),~~(i,j=1,2),
\end{eqnarray}
where $\Sigma^{\tilde{t}\prime}_{ij}(m_{\tilde{t}_j}^2)\equiv
\frac{\partial \Sigma^{\tilde{t}}_{ij}(p^2) }{\partial
p^2}|_{p^2=m_{\tilde{t}_j}^2}$, $\Sigma^{\tilde{t}}_{ij}\sim
{{\cal{O}}(\alpha_s)}$ and
$(\mu_{\tilde{t}_j}^2-m_{\tilde{t}_j}^2)\sim {{\cal{O}}(\alpha_s)}$.
We neglect the higher order terms and get approximately the mass and
wave function renormalization counterterms of top squarks as
\begin{eqnarray}
\label{eq4-1} \delta \mu^2_{\tilde{t}_i}  &=&
\Sigma_{ii}^{\tilde{t}}(m_{\tilde{t}_i}^2) +
(\mu^2_{\tilde{t}_i}-m_{\tilde{t}_i}^2)
\Sigma_{ii}^{\tilde{t}\prime}(m_{\tilde{t}_i}^2) ,~~ \\
\label{eq4-2} \delta Z_{ii}^{\tilde{t}} &=& -
\Sigma_{ii}^{\tilde{t}'}(m_{\tilde{t}_i}^2) ,   \\
\label{eq4-3}\delta Z_{ij}^{\tilde{t}} &=& \frac{2
\Sigma_{ij}^{\tilde{t}}(m_{\tilde{t}_j}^2)}
{m_{\tilde{t}_i}^2-m_{\tilde{t}_j}^2},~~(~i,j=1,2, i\neq j).
\end{eqnarray}

\par
By adopting the unitary condition for the bare and renormalized
mixing matrices of the top squark sector, $U^{\tilde{t}}_0$ and
$U^{\tilde{t}}$, we obtain the expression for the counterterm of top
squark mixing matrix $\delta U^{\tilde{t}}$ as
\begin{eqnarray}\label{eq25}
\delta U^{\tilde t} &=&\frac{1}{4} \left(\delta Z^{\tilde{t}}-\delta
Z^{\tilde{t}\dagger} \right)U^{\tilde{t}}.
\end{eqnarray}
With the definitions shown in Eqs.(\ref{eq-dectsf}) and
(\ref{eq-dectU}), the counterterms of the top squark mixing matrix
elements can be written as
\begin{eqnarray}\label{eq-26}
\delta U^{\tilde t}_{ij} &=& \frac{1}{4} \sum_{k=1}^2\left(\delta
Z^{\tilde{t}}_{ik} - \delta Z^{\tilde{t}\ast}_{ki}
\right)U^{\tilde{t}}_{kj} ~~(i,j=1,2).
\end{eqnarray}
The explicit expressions for unrenormalized self-energies of top
squarks, $\Sigma_{ij}^{\tilde{t}}(p^2)~(i,j=1,2)$, are written as
\begin{eqnarray}\label{self-1}
\Sigma_{ii}^{\tilde{t}}(p^2)&=&-\frac{\alpha_s}{3\pi}\left( 4
A_0[m_t^2]+A_0[\mu_{\tilde{t}_i}^2]\right)~~\nonumber \\
&+&\frac{4\alpha_s}{3\pi}\left[m_t
m_{\tilde{g}}(U_{i2}^{\tilde{t}}U_{i1}^{\tilde{t}\ast}
+U_{i1}^{\tilde{t}}U_{i2}^{\tilde{t}\ast})-m_{\tilde{g}}^2\right]
B_0[p^2,m_{\tilde{g}}^2,m_t^2]~~\nonumber \\
&-&\frac{4\alpha_s}{3\pi}p^2 \left(B_0[p^2,0,\mu^2_{\tilde{t}_i}]
+B_1[p^2,0,\mu^2_{\tilde{t}_i}]
+B_1[p^2,m_{\tilde{g}}^2,m_t^2] \right)~~\nonumber \\
&+&\frac{\alpha_s}{3\pi}\sum_{k=1,2}(U_{k1}^{\tilde{t}}U_{i1}^{\tilde{t}\ast}
-U_{k2}^{\tilde{t}}U_{i2}^{\tilde{t}\ast})(U_{i1}^{\tilde{t}}U_{k1}^{\tilde{t}\ast}
-U_{i2}^{\tilde{t}}U_{k2}^{\tilde{t}\ast})A_0[\mu_{\tilde{t}_k}^2],
\end{eqnarray}
\begin{eqnarray}\label{self-2}
\Sigma_{ij}^{\tilde{t}}(p^2)&=&\frac{4\alpha_s}{3\pi}m_t
m_{\tilde{g}}(U_{i2}^{\tilde{t}}U_{j1}^{\tilde{t}\ast}
+U_{i1}^{\tilde{t}}U_{j2}^{\tilde{t}\ast})B_0[p^2,m_{\tilde{g}}^2,m_t^2]~~\nonumber \\
&+&\frac{2\alpha_s}{3\pi}\sum_{k=1,2}(U_{k1}^{\tilde{t}}U_{j1}^{\tilde{t}\ast}
-U_{k2}^{\tilde{t}}U_{j2}^{\tilde{t}\ast})(U_{i1}^{\tilde{t}}U_{k1}^{\tilde{t}\ast}
-U_{i2}^{\tilde{t}}U_{k2}^{\tilde{t}\ast})A_0[\mu_{\tilde{t}_k}^2],\nb \\
&&~~~~~~~~~~~~~~~~~~~~~~(i\neq j,~i,j=1,2).
\end{eqnarray}
From Eqs.(\ref{eq4-2}), (\ref{eq4-3}), (\ref{self-1}) and
(\ref{self-2}), we can obtain $Im(\delta Z_{ij}^{\ast})=Im(\delta
Z_{ji})$. Equation (\ref{eq-26}) tells us that we can choose both the
renormalized mixing matrix elements $U^{\tilde{t}}$ and their
counterterms $\delta U^{\tilde{t}}$ made up of real matrix elements.
Then the matrix $U^{\tilde{t}}$ can be taken explicitly as
\begin{eqnarray}
U^{\tilde t} &=& \left(\begin{array}{cc} \cos\theta_t &\sin\theta_t   \\
-\sin\theta_t & \cos\theta_t
\end{array} \right).
\end{eqnarray}

\par
With the Lagrangian shown in Eq.(\ref{Lagrangian-2}), the
counterterms of the $\tilde t_i - \bar d_k - e$, $\tilde d^*_{k,i} -
t - e$ and $\tilde e - \bar d_k - t$ vertexes are expressed
correspondingly as below:
\begin{eqnarray}
\label{eq-Vst} &&\delta V_{\tilde t_i \bar d_k e}
=-\lambda^\prime_{13k}\left(\frac{\delta\lambda^\prime_{13k}}{\lambda^\prime_{13k}}
+\frac{\delta U^{\tilde t}_{i1}}{U^{\tilde t}_{i1}}+
\frac{1}{2}\delta Z_R^{d_k} + \frac{1}{2}\frac{U^{\tilde
t}_{11}}{U^{\tilde t}_{i1}}\delta Z^{\tilde t}_{1i} +
\frac{1}{2}\frac{U^{\tilde t}_{21}}{U^{\tilde t}_{i1}}\delta Z^{\tilde t}_{2i} \right),~~(i=1,2),\\
\label{eq-Vsd} &&\delta V_{\tilde d^*_{k,i} t e}
=-\lambda^\prime_{13k}\left(\frac{\delta\lambda^\prime_{13k}}{\lambda^\prime_{13k}}
+\frac{\delta U^{\tilde d}_{i1}}{U^{\tilde d}_{i1}}+
\frac{1}{2}\delta Z_L^{t} + \frac{1}{2}\frac{U^{\tilde
d}_{11}}{U^{\tilde d}_{i1}}\delta Z^{\tilde d}_{1i} +
\frac{1}{2}\frac{U^{\tilde d}_{21}}{U^{\tilde d}_{i1}}\delta Z^{\tilde d}_{2i} \right),~~(i=1,2),\\
\label{eq-Vse} &&\delta V_{\tilde e \bar d_k t}
=-\lambda^\prime_{13k}\left(\frac{\delta\lambda^\prime_{13k}}{\lambda^\prime_{13k}}
+\frac{1}{2}\delta Z_R^{d_k}+\frac{1}{2}\delta Z_L^t\right),
\end{eqnarray}
where the lower index $k(=1,2)$ denotes the quark/squark generation.
In this work we do not adopt the decoupling scheme described in
Ref.\cite{Decouple} as we do not consider the scenarios where the
renormalization and factorization scales are much larger than the
top squark masses. By using the $\overline{MS}$ scheme to
renormalize the $\lambda^{\prime}_{13k}$ coupling, we get
\begin{equation}
\delta\lambda^\prime_{13k}=\lambda^\prime_{13k}\left(-\frac{C_F}{2}\right)
\frac{\alpha_s}{\pi\bar{\epsilon}},~~(k=1,2),
\end{equation}
where $1/\bar\epsilon=1/\epsilon_{UV}-\gamma_E+ln(4\pi)$ and
$C_F=4/3$. After the renormalization procedure we get a UV-finite
virtual correction to the partonic process $e^+ d_k\to t\chi^0_1$.

\par
\begin{figure*}
\begin{center}
\includegraphics[ scale = 0.75 ]{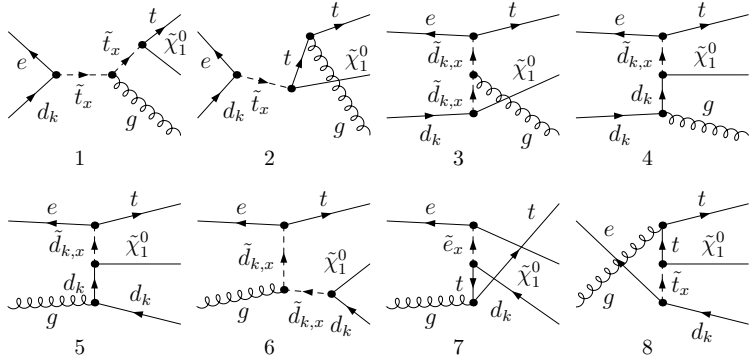}
\caption{\label{fig3} Some representative Feynman diagrams for the
real gluon (1-4) and light-quark (5-8) emission processes. }
\end{center}
\end{figure*}

According to the Kinoshita-Lee-Nauenberg(KLN) theorem
\cite{KLN_1,KLN_2}, the contributions of the real gluon emission
process, $e^+(p_1) d_k(p_2)\to t(p_3)\chi^0_1(p_4)g(p_5)$, and the
light-quark emission process, $e^+(p_1)g (p_2)\to
t(p_3)\chi^0_1(p_4)\bar d_k(p_5)$, are at the same $\alpha_s$ order
as the virtual correction to the partonic process $e^+ d_k\to
t\chi^0_1$ in perturbative calculation. Part of the tree-level
Feynman diagrams for those two processes are depicted in
Fig.\ref{fig3}. We adopt the two-cutoff phase space slicing 
(TCPSS) method \cite{TCPSS} to isolate the IR divergence of the real
emission processes and make a cross-check with the result by using
the dipole subtraction (DPS) method \cite{dipole}.

\par
In the TCPSS method, two arbitrary small cutoff parameters, soft
cutoff $\delta_s$ and collinear cutoff $\delta_c$, are introduced
for the real emission process. The phase space of the real gluon
emission process is divided into two regions: the soft region
($E_{5}\leq \delta_{s}\sqrt{\hat{s}}/2$) and the hard region
($E_{5}>\delta_{s}\sqrt{\hat{s}}/2$). $\delta_c$ separates the hard
region into the hard collinear (HC) region where $\hat{s}_{25} <
\delta_{c}\hat{s}$, and hard noncollinear ($\overline {HC}$)
region. Then the cross section for the real gluon emission partonic
process can be written as
\begin{equation}\label{eq-realgluon}
\hat \sigma^{R}_{g} ( e^+d_k \to t\tilde\chi^0_1g ) = \hat
\sigma^{S}_{g}+\hat \sigma^{H}_{g} = \hat \sigma^{S}_{g}+ \hat
\sigma^{HC}_{g}+\hat \sigma^{\overline {HC}}_{g}.
\end{equation}
For the real light-quark emission process, the phase space is split
into collinear (C) and noncollinear ($\overline C$) regions by
cutoff $\delta_c$, and the cross section for $e^+(p_1)g (p_2)\to
t(p_3)\tilde\chi^0_1(p_4)\bar d_k(p_5)$ is obtained as
\begin{equation}\label{eq-lightquark}
\hat \sigma^{R}_{q} ( e^+g \to t\tilde\chi^0_1\bar d_k ) = \hat
\sigma^{C}_{q}+ \hat\sigma^{\overline C}_{q}.
\end{equation}
There is no divergence in the noncollinear region, so
$\hat{\sigma}_g^{\overline{HC}}$ in Eq.(\ref{eq-realgluon}) and
$\hat{\sigma}_q^{\overline{C}}$ in Eq.(\ref{eq-lightquark}) are
finite and can be evaluated in four dimensions by using the general
Monte Carlo method. After summing all the contributions mentioned
above there still exists the remained collinear divergence, which
will be absorbed by the redefinition of the PDFs at the NLO.

\par
The total NLO QCD correction includes the virtual correction
$\sigma^V$, real gluon emission correction $\sigma^R_g$ and
light-quark emission correction $\sigma^R_q$. The full NLO QCD
correction to the cross section for the $e^+p\to t\tilde\chi^0_1+X$
process at the LHeC is formally given by the QCD factorization
formula as
\begin{eqnarray}\label{eq-NLOcs}
\Delta\sigma_{NLO}&=& \int_0^1 dx \left\{ G_{d_k/P}(x,\mu_f)
\left[\hat{\sigma}^V(xs) + \hat\sigma^S_g(xs) +
\hat\sigma^{\overline{HC}}_g(xs)\right] \right. \nonumber \\
&&\left. + G_{g/P}(x,\mu_f) \hat\sigma^{\overline C}_q(xs) +
G^\prime_{d_k/P}(x,\mu_f)\hat\sigma^{HC}_g(xs)
+ G^\prime_{g/P}(x,\mu_f) \hat\sigma^{C}_q(xs)\right\} \nonumber \\
&=&\sigma^V+\sigma^S_g+\sigma^{HC}_g+\sigma^{C}_q+\sigma^{\overline{HC}}_{g}+
\sigma^{\overline C}_{q},
\end{eqnarray}
where $G_{d_k(g)/P}(x,\mu_f)$ denote the PDFs and
$G^\prime_{d_k(g)/P}(x,\mu_f)$ are defined as \cite{TCPSS}
\begin{equation}
G^\prime_{q/P}(x,\mu_f)=\int_{x}^{1-\delta_s}\frac{dz}{z}G_{q/P}(x/z,\mu_f)P_{qq},~~
G^\prime_{g/P}(x,\mu_f)=\int_{x}^{1}\frac{dz}{z}G_{g/P}(x/z,\mu_f)P_{qg},
\end{equation}
where
\begin{equation}
P_{qq}=\frac{4}{3}\frac{1+z^2}{1-z}\ln\left(\delta_c\frac{1-z}{z}
\frac{\hat s}{\mu^2_f}\right)+\frac{4}{3}(1-z),~~
P_{qg}=\frac{1}{2}\left[(1-z)^2+z^2\right]\ln\left(\delta_c\frac{1-z}{z}
\frac{\hat s}{\mu^2_f}\right)+z(1-z).
\end{equation}
The NLO QCD correction $\Delta\sigma_{NLO}$ can be divided into the 
two-body term $\Delta\sigma^{(2)}$ and three-body term
$\Delta\sigma^{(3)}$. The two-body term is defined as
$\Delta\sigma^{(2)}=\sigma^V+\sigma^S_g+\sigma^{HC}_{g}+\sigma^C_{q}$.
The three-body term consists of two contribution parts, i.e.,
$\Delta\sigma^{(3)}=\sigma^{\overline {HC}}_g +\sigma^{\overline
C}_{q}$. Finally, the QCD corrected total cross section for the
$e^+p \to t\tilde\chi^0_1+X$ process is written as $\sigma_{NLO}
=\sigma_{LO} +\Delta \sigma^{(2)} +\Delta \sigma^{(3)}= \sigma_{LO}
+ \Delta\sigma_{NLO}$.

\par
Analogously, we can evaluate the LO and NLO QCD corrected integrated
cross sections for the $e^-p\to \bar t\tilde\chi^0_1+X$ process at
the LHeC.

\vskip 5mm
\section{Numerical results and discussion }
\par
In this section we present and discuss the numerical results for the
$e^+p(e^-p)\to t\chi^0_1(\bar t\chi^0_1) +X$ process at the LHeC
with $E_e=50(150)~GeV$ and $E_p=7~TeV$ in proton-electron (positron)
collision mode. In our calculation, we take one-loop and two-loop
running $\alpha_s$ in the LO and NLO calculations, respectively
\cite{Amsler}, and the number of active flavors is taken as $N_f=4$.
The QCD parameter $\Lambda_4^{LO}=215~MeV$ and the CTEQ6L1 PDFs are
adopted in the LO calculation, while
$\Lambda_4^{\overline{MS}}=326~MeV$ and the CTEQ6M PDFs are used in
the NLO calculation \cite{cteq_1,cteq_2}. We set the factorization scale
and renormalization scale to be equal, and take $\mu\equiv\mu_0=
m_{\tilde t_1}$ in default. The Cabibbo-Kobayashi-Maskawa matrix is set to the unit
matrix. We ignore the masses of electrons and $u$-, $d$-, $s$-, and
$c$-quarks, and take the related SM parameters as $m_t=173.5~GeV$,
$m_W=80.385~GeV$, $m_Z=91.1876~GeV$ and $\alpha_{ew}=1/137.036$ in
numerical calculation \cite{Amsler,pdglive}.

\par
In analyzing the $e^- p \to \bar{t} \tilde{\chi}_1^0 + X$ and $e^+ p
\to t \tilde{\chi}_1^0 + X$ processes at the LHeC, we adopt the
single dominance hypothesis \cite{superpotential_3} that all the
$R$-violating couplings are zero except one typical $R$-violating
coupling. In this paper, two cases of $R$-violating parameter values
are considered. Case (1): All the $R$-violating couplings are zero
except $\lambda_{131}^{\prime} = 0.1$; case (2):
$\lambda_{132}^{\prime} = 0.1$ and other $R$-violating couplings are
zero.

\par
There are many discussions about the LHC experimental constraints on
the SUSY parameters, For recent papers, see
Refs.\cite{higgsfit,sigmis1,conshiggs} for the data analysis
concerning the $\sim 125~GeV$ Higgs-like boson, Ref.\cite{consDM}
for dark matter, Ref.\cite{consfo} for low energy flavor observables
and so on. In our work, we choose a benchmark point in the pMSSM-19
for numerical demonstration. Considering recent constraints from the
ATLAS and CMS experiments \cite{CMS-A,Atlas-A,Atlas-B,atlas,Feng}
and the ''Review of Particle Physics'' in Ref.\cite{pdglive}, we
take the 19 free input parameters as listed below.
\begin{eqnarray}\label{para-susy1}
&&\mu=231.21 ~GeV, ~~\tan\beta=32.02, ~~M_1=1633.48
~GeV,~~M_2=143.01 ~GeV, \nonumber\\
&& M_3=1500~GeV,
~~A_{t}=1734.17~GeV,~~ A_b=-1150.85~GeV,~~A_{\tau}=-1879.01~GeV, \nonumber\\
&&m_{A^0}=2620.14~GeV,~~m_{\tilde e_L}=m_{\tilde \mu_L}=290~GeV,~~
m_{\tilde e_R}=m_{\tilde \mu_R}=550~GeV, ~~m_{\tilde \tau_L}=2612.45~GeV, \nonumber\\
&&m_{\tilde \tau_R}=2680.28~GeV,~~m_{\tilde q_{1L}}=m_{\tilde q_{2L}}=1825.87~GeV,~~
m_{\tilde q_{3L}}=790~GeV,  \nonumber\\
&&m_{\tilde u_R}=m_{\tilde c_R}=2035.93~GeV,~~m_{\tilde t_R}=1115.39~GeV,  \nonumber\\
&&m_{\tilde d_R}=m_{\tilde s_R}=3162.04~GeV,~~m_{\tilde
b_R}=1533.0~GeV.
\end{eqnarray}
After applying the modified {\small SOFTSUSY} 3.3.7 program \cite{softsusy}
with above 19 input parameters in the RPV pMSSM-19 with the 
$\lambda_{131}^{\prime} = 0.1$ (or $\lambda_{132}^{\prime} = 0.1$)
and other $R$-violating couplings being zero, we get the SUSY
benchmark point with the parameters related to our calculations as
\begin{eqnarray}\label{para-susy2}
&&m_{\tilde t_1}=758.81~GeV, ~~m_{\tilde t_2}=1142.60 ~GeV, ~~
\theta_{\tilde t}=-21.546^\circ, ~~
m_{\tilde b_1}=813.15~GeV, \nonumber\\
&&m_{\tilde b_2}=1544.27~GeV,~~\theta_{\tilde b}=0.5704^\circ,~~
m_{\tilde e_1}=283.70~GeV, ~~m_{\tilde e_2}=563.93~GeV, \nonumber\\
&&m_{\tilde g}=1586.36~GeV,~~m_{\tilde u_R}=m_{\tilde c_R}=2052.18~GeV,~~
m_{\tilde u_L}=m_{\tilde c_L}=1849.96~GeV, \nonumber\\
&&m_{\tilde d_R}=m_{\tilde s_R}=3177.58~GeV,~~m_{\tilde
d_L}=m_{\tilde s_L}=1851.45~GeV,~~m_{\tilde\chi^0_1}=128.59~GeV, \nonumber\\
&&m_h=125.48~GeV,~~m_H=m_A=m_{H^{\pm}}=2.62~TeV.
\end{eqnarray}
As we know that the top squarks can decay to $R$-even particles via
the nonzero $R$-parity violating couplings in the RPV pMSSM-19,
i.e., $\tilde{t}_{1,2} \to e^+ d$ and $\tilde{t}_{1,2} \to e^+ s$ in
case (1) and case (2), respectively. The corresponding RPV decay
widths are expressed as
\begin{equation}\label{eqn:G-RPVstop}
\Gamma_{\tilde{t}_i}^{RPV} =
|U^{\tilde{t}}_{i1}|^2 \frac{{\lambda_{13k}^{'2}}}{16\pi } \cdot
\frac{{(m_{\tilde{t}_i}^{2}-m_{d_k}^{2})}^{2}}{m_{\tilde{t}_i}
{(m_{\tilde{t}_i}^{2}+m_{d_k}^{2})}},~~(i,k=1,2).
\end{equation}
With the SUSY parameters in Eq.(\ref{para-susy2}) we obtain the
$\tilde{t}_{1}$ and $\tilde{t}_{2}$ $R$-parity conservation (RPC)
decay widths as $\Gamma_{\tilde{t}_1}^{RPC}=18.29 ~GeV$,
$\Gamma_{\tilde{t}_2}^{RPC}=58.10 ~GeV$ by applying the {\small SUSY-HIT} 1.3
program \cite{susyhit}, and the RPV decay widths for $\tilde{t}_{1}$
and $\tilde{t}_{2}$ as $\Gamma_{\tilde{t}_1}^{RPV}= 0.131 ~GeV$,
$\Gamma_{\tilde{t}_2}^{RPV}=0.031~GeV$ from Eq.(\ref{eqn:G-RPVstop})
by using our developed program. Then the total decay widths for top
squarks are
$\Gamma_{\tilde{t}_1}=\Gamma_{\tilde{t}_1}^{RPC}+\Gamma_{\tilde{t}_1}^{RPV}=18.42~GeV$
and
$\Gamma_{\tilde{t}_2}=\Gamma_{\tilde{t}_2}^{RPC}+\Gamma_{\tilde{t}_2}^{RPV}=58.13~GeV$£¬
respectively.

\par
We can see from Eq.(\ref{para-susy2}) that the mass value of the
lightest neutral Higgs boson is coincident with the recent LHC
result on the discovery of a $125~GeV$ Higgs boson, and the masses
for remaining Higgs particles escape from the experimental exclusion
constraints \cite{Atlas-A,Atlas-C}. We compare the various signal
strengths obtained by adopting our chosen pMSSM-19 parameter set
with those from recent CMS experimental reports for the $125~GeV$
Higgs boson \cite{CMS-A}, and find that they are compatible. With
the SUSY parameters in Eq.(\ref{para-susy2}) we obtain the signal
strengths, defined as $\mu=\sigma/\sigma_{SM}$, for $125~GeV$ Higgs
as $\mu(h \to ZZ^{(*)})=1.0349$, $\mu(h \to \gamma\gamma)=1.044$,
$\mu(h \to WW^{(*)})=0.9998$, $\mu(h \to b\bar b)=0.9599$, $\mu(h
\to \tau\tau)=1.0181$ and the combined signal strength $\mu=0.9772$.
While the corresponding recent CMS experimental data for $m_h =
125.7~GeV$ from Ref.\cite{CMS-A} are as: $\mu(h \to
ZZ^{(*)})=0.92\pm 0.28$, $\mu(h \to\gamma\gamma)=0.77 \pm 0.27$,
$\mu(h \to WW^{(*)})=0.68 \pm 0.20$, $\mu(h \to b\bar b)=1.15\pm
0.62$, $\mu(h \to \tau\tau)=1.10\pm 0.41$ and the combined signal
strength $\mu=0.80 \pm 0.14$. We can see that they are in agreement
with each other within measurement errors except for $\mu(h \to
WW^{(*)})$ whose theoretical value is approximately coincident with
that in Ref.\cite{CMS-A}, but fitted nicely with the ATLAS result in
Ref.\cite{Atlas-B} (i.e., $\mu(h \to WW^{(*)})=1.0 \pm 0.3$).

\par
In the RPV pMSSM-19 model the lightest supersymmetric particle
(LSP), $\tilde{\chi}_1^0$, is no longer stable. It will decay into
$R$-even particles via the $R$-violating interactions. That means
$\tilde{\chi}_1^0$ decays into $bd\nu_{e}$ and $bs\nu_{e}$ in case
(1) and case (2), respectively. Then the typical reaction chains are
\begin{eqnarray} \label{epchain}
&&e^-p \to \bar{t}\tilde{\chi}_1^0+ X \to \bar{t}(bd\nu_e)+X \to
(\bar{b}W^-)(bd\nu_e)+X,
~~for~Case~(1),  \nonumber\\
&&e^+p \to t\tilde{\chi}_1^0+ X \to t(bs\nu_e)+X \to
(bW^+)(bs\nu_e)+X, ~~for~Case~(2).
\end{eqnarray}
If we choose the detection of $W$-bosons by means of the decay $W\to
l\nu$, the typical signatures for the above processes would be
$2~b$-$jets$+$jet$+$l$+$\slash\hspace{-8pt}E_T$ \cite{hera1}.

\par
In order to demonstrate the necessity for calculating the complete
Feynman diagrams shown in Figs.\ref{fig1}(a)-(c), in Table
\ref{tab-2} we present the numerical results for the tree-level
integrated cross sections by adopting three approaches
($\sigma^{NWA}_{LO}$, $\sigma^{s}_{LO}$, $\sigma^{All}_{LO}$): (1)
the narrow-width approximation (NWA) where we factorize the
production and decay of top squark, (2) the total cross section
contributed only by the s-channel diagrams [Fig.\ref{fig1}(a)], (3)
the total cross section contributed by complete s-, t-, u-channel
diagrams [Figs.\ref{fig1}(a)-(c)]. The relative discrepancies
$\delta_1$ and $\delta_2$ are defined as $\delta_1\equiv
(\sigma^{All}_{LO}-\sigma^{NWA}_{LO}) /\sigma^{All}_{LO}$ and
$\delta_2\equiv
(\sigma^{All}_{LO}-\sigma^{s}_{LO})/\sigma^{All}_{LO}$,
respectively. The numerical calculations are carried out with the
SUSY parameters in Eq.(\ref{para-susy2}), $E_e=50~GeV$, $E_p=7~TeV$
and the $R$-violating parameters being in case (1) for the processes
$e^+p \to t\tilde\chi^0_1+X$, $e^-p \to\bar{t}\tilde\chi^0_1+X$, and
in case (2) for the $e^+p \to t\tilde\chi^0_1+X$ process. From the
results in this table we can see that in our precision investigation
on these $R$-parity violating processes we should consider
completely all the relevant diagrams including the nonresonant
diagrams for these processes at the LHeC.
\begin{table}
\begin{center}
\begin{tabular}{|c|c|c|c|c|c|}
\hline Process  & $\sigma^{NWA}_{LO}$(fb) & $\sigma^{s}_{LO}$(fb)
& $\sigma^{All}_{LO}$(fb) & $\delta_1(\%)$ & $\delta_2(\%)$ \\
\hline \tabincell{c}{$e^+p \to t\tilde\chi^0_1+X$\\in Case (1)}
& 10.73 & 18.29 & 21.24 & 49.5 & 13.9 \\
\hline \tabincell{c}{$e^-p\to \bar t\tilde\chi^0_1+X$\\in Case(1)}
& 0.4101 & 0.7952 & 1.410 & 70.9 & 43.6 \\
\hline \tabincell{c}{$e^+p \to t\tilde\chi^0_1+X$ \\ in Case (2)}
& 0.2657 & 0.4366 & 0.7447 & 64.3 & 41.4 \\
\hline
\end{tabular}
\end{center}
\caption{ \label{tab-2} The numerical results of the LO cross
sections for the processes $e^+p \to t\tilde\chi^0_1+X$, $e^-p
\to\bar{t}\tilde\chi^0_1+X$ in case (1) and $e^+p \to
t\tilde\chi^0_1+X$ in case (2) at the benchmark point at the
$E_e=50~GeV$, $E_p=7~TeV$ LHeC by adopting three approaches.
$\sigma^{NWA}_{LO}$ is the LO cross section by applying the NWA.
$\sigma^{s}_{LO}$ is obtained by considering only the $s-$channel
diagram, and $\sigma^{All}_{LO}$ is contributed by all the $s-$, $t-$,
u-channel diagrams. $\delta_1$ and $\delta_2$ are defined as
relative discrepancies of the integrated cross sections. }
\end{table}
\par
\begin{figure}[htbp]
\includegraphics[width=0.5\textwidth]{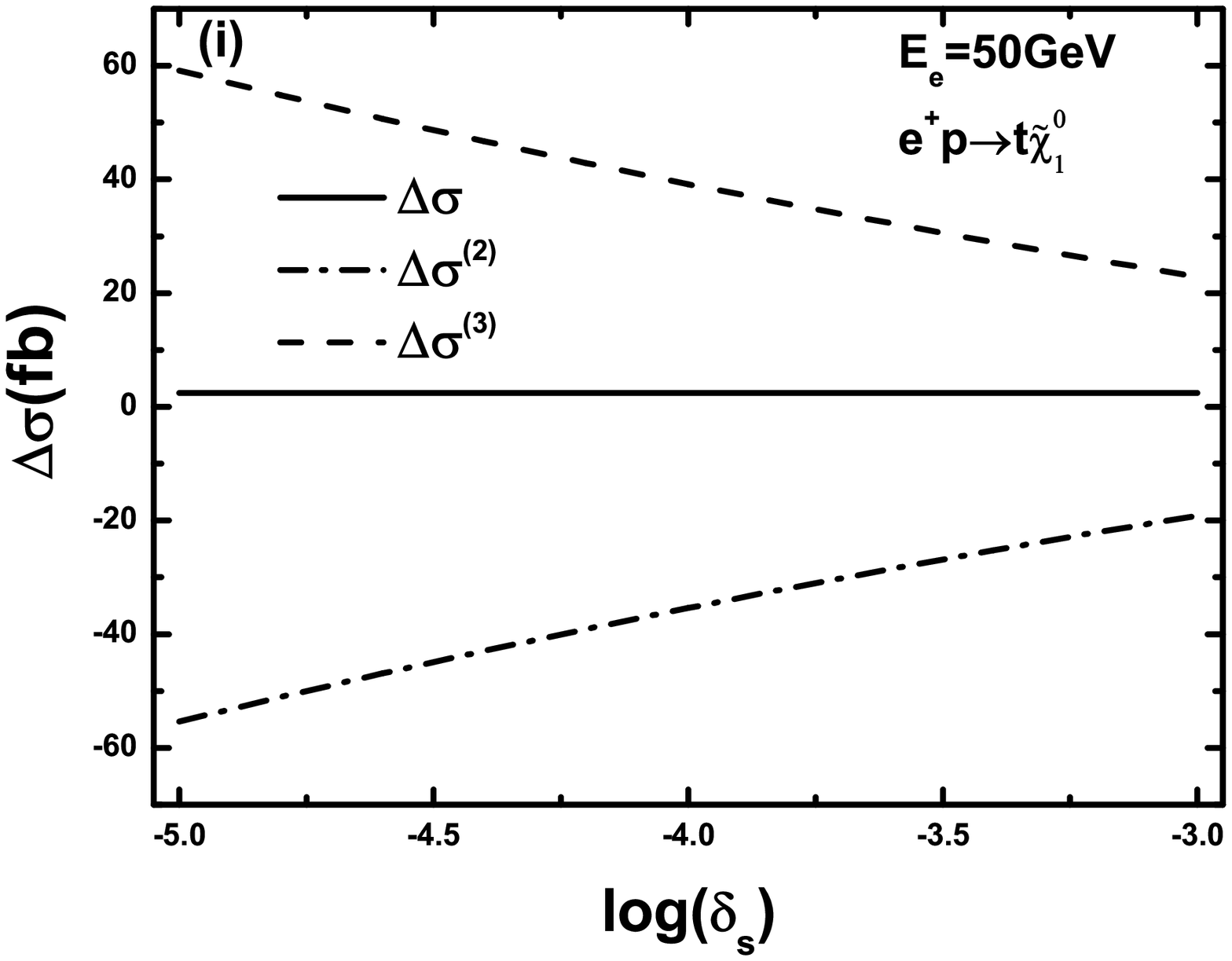}%
\hspace{0in}%
\includegraphics[width=0.5\textwidth]{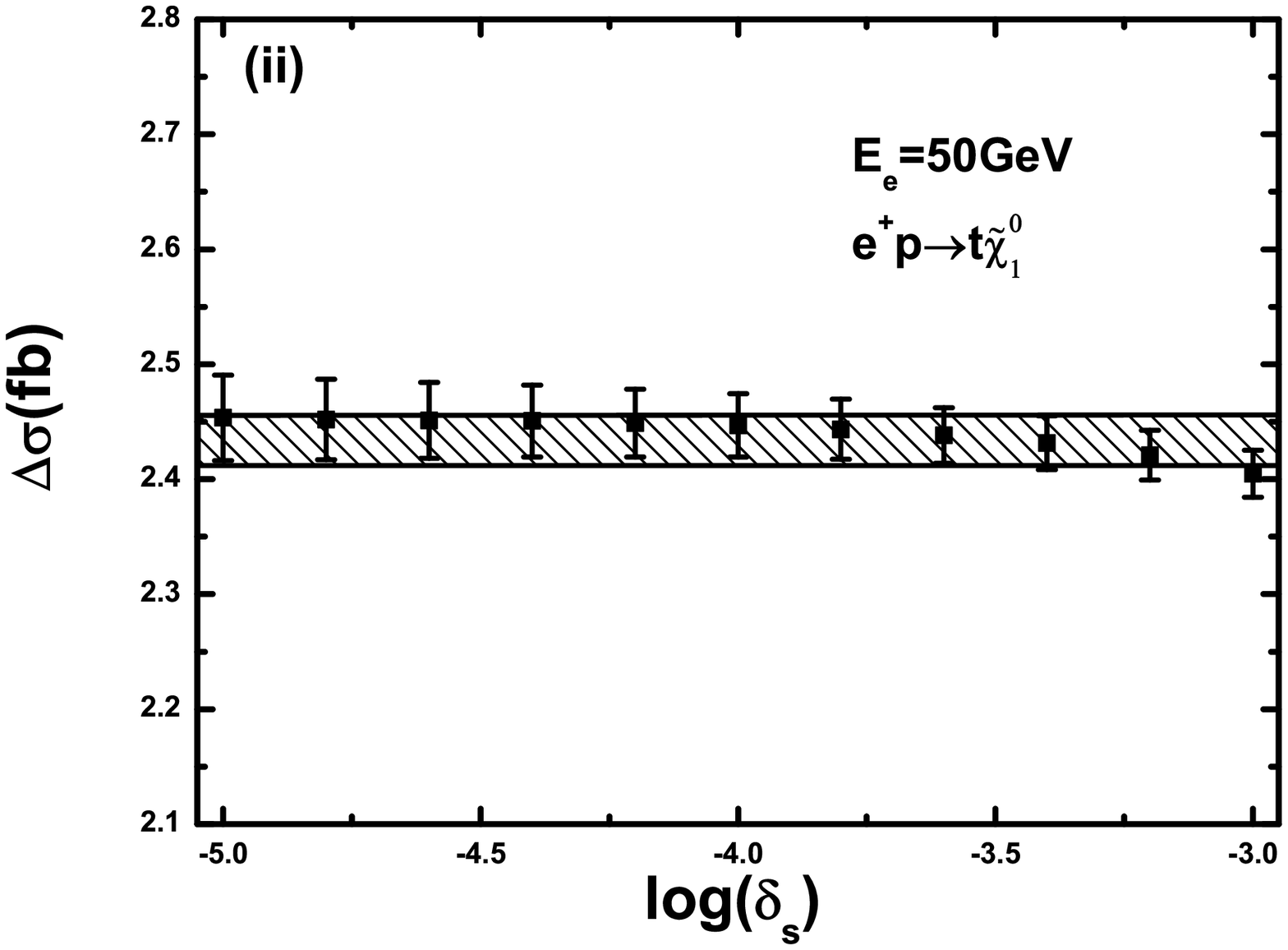}%
\hspace{0in}%
\caption{\label{fig-dels} (i) The dependence of the NLO QCD
corrections to the $e^+p\to t\tilde\chi^0_1 + X$ process on the soft
cutoff $\delta_s$ with the $R$-violating parameters in case (1) at
the $E_e=50~GeV$ and $E_p=7~TeV$ LHeC by taking $\delta_c =
\delta_s/50$. (ii) The NLO QCD correction $\Delta\sigma_{NLO}$ with
calculation error as the function of $\delta_s$. The result from the
DPS method with $\pm\sigma$ calculation error is shown in the
shadowed region. }
\end{figure}

Figure \ref{fig-dels} demonstrates that the total NLO QCD correction to
the $e^+p\to t\tilde\chi^0_1 + X$ process with the $R$-violating
parameters in case (1) at the $E_e=50~GeV$ and $E_p=7~TeV$ LHeC,
does not depend on the arbitrarily chosen values of $\delta_s$ and
$\delta_c$ within the calculation errors, where we take the SUSY
parameters shown in Eq.(\ref{para-susy2}),
$\lambda^{\prime}_{131}=0.1$ and other related $\lambda^{\prime}=0$.
In Fig.\ref{fig-dels}(a), we plot the two-body correction
$\Delta\sigma^{(2)}$, the three-body correction $\Delta\sigma^{(3)}$,
and the total QCD correction $\Delta\sigma_{NLO}$ for the $e^+p \to
t\tilde\chi^0_1 + X$ process as the functions of the soft cutoff
$\delta_s$, running from $1\times 10^{-5}$ to $1\times 10^{-3}$ with
$\delta_c=\delta_s/50$. In Fig.\ref{fig-dels}(b), the results for
$\Delta\sigma_{NLO}$ with calculation errors are depicted. We can
see that the total NLO QCD correction to the $e^+p \to
t\tilde\chi^0_1 + X$ process is independent of the arbitrary cutoff
$\delta_s$ and $\delta_c$ within the statistic errors. To make a
further verification of the correctness of the TCPSS results, we
adopt the DPS method by using the MadDipole
program \cite{maddipole,maddipole2} to deal with the IR
singularities. We present the DPS result with $\pm\sigma$
calculation error in the shadowed region shown
Fig.\ref{fig-dels}(b). It shows that all the results from both two
methods are in good agreement. In further numerical calculations, we
adopt the TCPSS method and take $\delta_s=5 \times 10^{-5}$ and
$\delta_c=\delta_s/50$.

\par
\begin{figure}[htbp]
\includegraphics[scale=0.4]{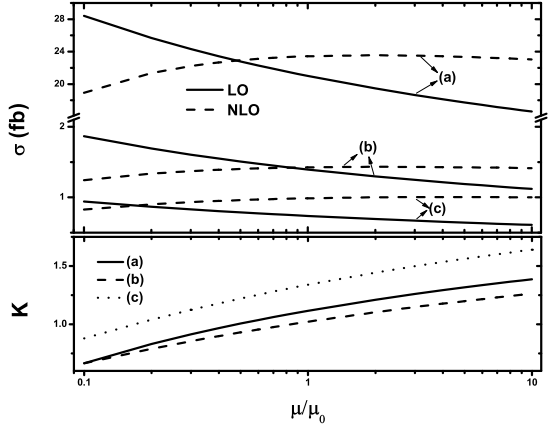}%
\hspace{0in}%
\caption{\label{figmu} The LO, NLO QCD corrected cross sections and
the corresponding $K$-factors versus the
factorization/renormalization scale ($\mu/\mu_0$) at the
$E_e=50~GeV$ and $E_p=7~TeV$ LHeC. The curves labeled with (a),
(b), and (c) indicate for the processes $e^+p \to
t\tilde\chi^0_1+X$, $e^-p \to\bar{t}\tilde\chi^0_1+X$ in case (1)
and $e^+p \to t\tilde\chi^0_1+X$ in case (2), respectively.  }
\end{figure}

Because the luminosities of $s$- and $\bar{s}$-quarks in proton are
the same, the observables for the process $e^-p\to e^-\bar s\to\bar
t\tilde\chi^0_1+X$ with the $R$-violating parameters in case (2) are
equal to the corresponding ones for the process $e^+p\to e^+s \to
t\tilde\chi^0_1+X$; we present only the plots for the $e^+p \to
t\tilde\chi^0_1+X$ process with the $R$-violating parameters in case
(2) in this paper. In Fig.\ref{figmu} we show the dependence of the
LO and the NLO QCD corrected cross sections for the processes $e^+p
\to t\tilde\chi^0_1+X$, $e^-p \to\bar{t}\tilde\chi^0_1+X$ with the
$R$-violating parameters in case (1), the and $e^+p\to
t\tilde\chi^0_1+X$ process with the $R$-violating parameters in case
(2) on the factorization/renormalization scale ($\mu/\mu_0$) at the
$E_e=50~GeV$ and $E_p=7~TeV$ LHeC by taking the SUSY parameters as
declared above. The corresponding $K$-factors
$(K\equiv\sigma_{NLO}/\sigma_{LO})$ are shown in the lower plot of
Fig.\ref{figmu}. In the figures, the curves labeled with (a),
(b), and (c) indicate the processes $e^+p \to
t\tilde\chi^0_1+X$, $e^-p\to \bar{t}\tilde\chi^0_1+X$ in case (1)
and $e^+p \to t\tilde\chi^0_1+X$ in case (2), respectively. There we
define $\mu_0\equiv m_{\tilde{t}_1}$ and set $\mu=\mu_r=\mu_f$ for
simplicity. The curves in Fig.\ref{figmu} for the processes $e^+p
\to t\tilde\chi^0_1+X$, $e^-p \to\bar{t}\tilde\chi^0_1+X$ in case
(1) and $e^+p \to t\tilde\chi^0_1+X$ in case (2), show that the NLO
QCD correction can reduce the factorization/renormalization scale
uncertainty of the integrated cross section. In the following
calculations we fix $\mu=\mu_0$.

\par
\begin{figure}[htbp]
\includegraphics[scale=0.4]{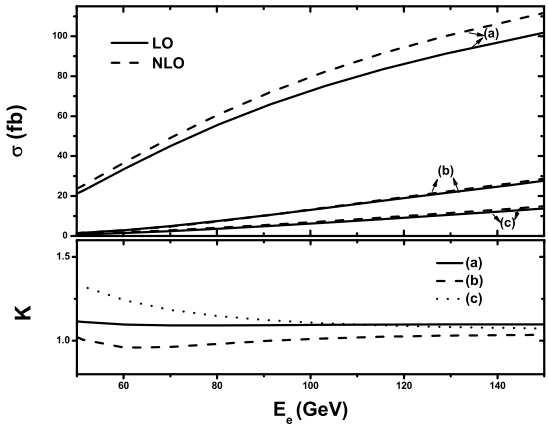}%
\hspace{0in}%
\caption{\label{fig6} The LO, NLO QCD corrected cross sections and
the corresponding $K$-factors versus the electron (positron) beam
energy $E_e$ at the LHeC. The curves labeled with (a), (b), and
(c) are for the processes $e^+p \to t\tilde\chi^0_1+X$, $e^-p\to
\to\bar{t}\tilde\chi^0_1+X$ in case (1) and $e^+p \to
t\tilde\chi^0_1+X$ in case (2), respectively. }
\end{figure}

The LO, NLO QCD corrected cross sections and the corresponding
$K$-factors versus electron (positron) beam energy $E_e$ with proton
beam energy $E_p=7~TeV$ at the LHeC at the benchmark point with SUSY
parameters shown in Eq.(\ref{para-susy2}) are depicted in
Fig.\ref{fig6}, respectively. The $R$-parity violating coupling
coefficients are set to be the values in case (1) for the LO and NLO
curves labeled with (a) and (b) which correspond to the processes
$e^+p \to t\tilde\chi^0_1+X$ and $e^-p \to\bar{t}\tilde\chi^0_1+X$
,respectively, while for the curve (c), which corresponds to process
$e^+p \to t\tilde\chi^0_1+X$ we take the $R$-violating parameters as
in case (2). We can see that the NLO QCD corrections for the process
$e^+p \to t\tilde\chi^0_1+X$ in case (1) and the process $e^+p \to
t\tilde\chi^0_1+X$ in case (2) enhance the LO results, respectively,
while the NLO correction for the $e^-p \to\bar{t}\tilde\chi^0_1+X$
process in case (1) suppresses the LO cross section in the range of
$55~GeV < E_e < 90~GeV$.

\par
\begin{figure}[htbp]
\includegraphics[scale=0.4]{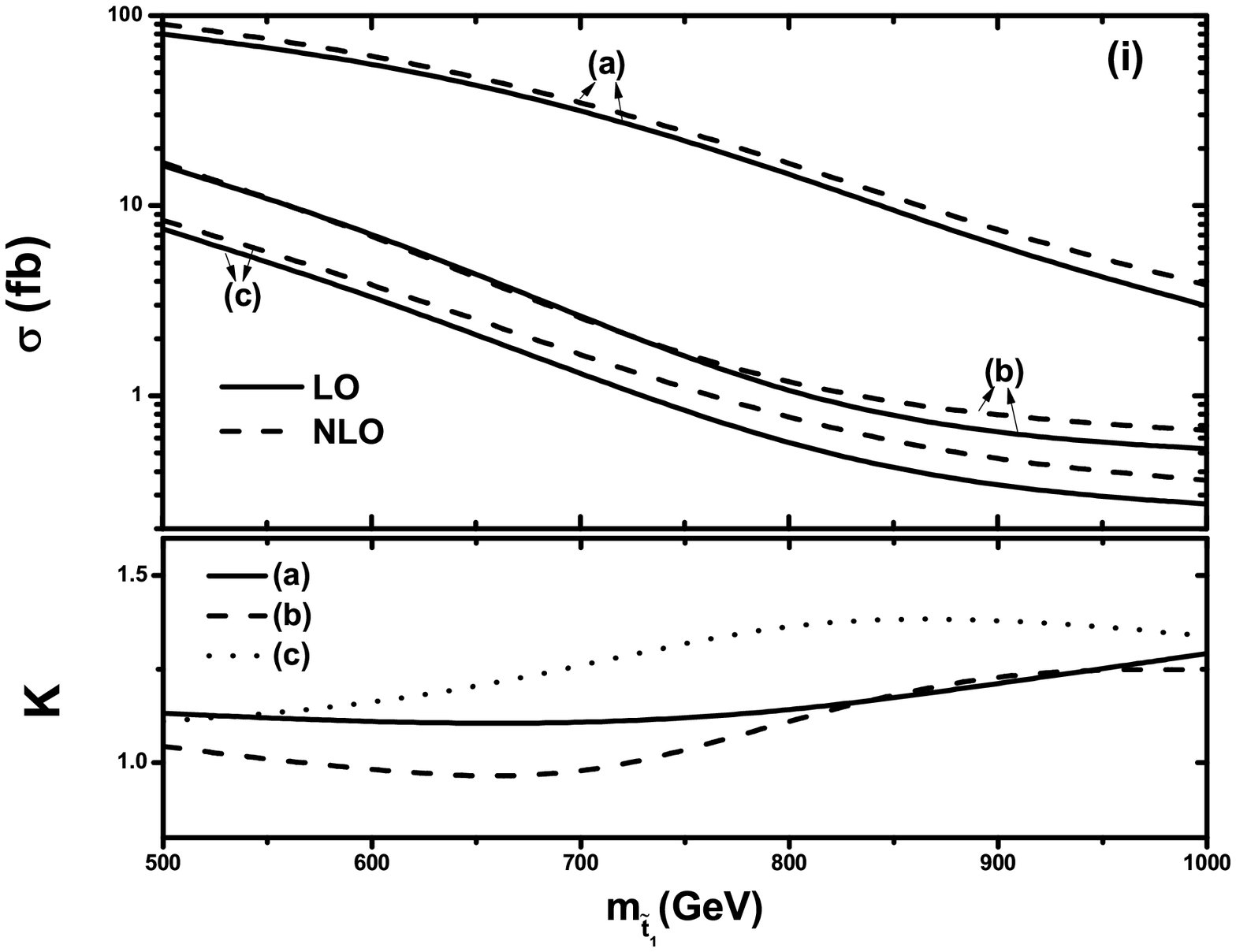}%
\hspace{0in}%
\caption{\label{fig7} The LO, NLO QCD corrected cross sections and
the corresponding $K$-factors versus the mass of top squark
$m_{\tilde t_1}$ at the $E_e=50~GeV$ and $E_p=7~TeV$ LHeC. The
curves labeled with (a), (b), and (c) indicate for the
processes $e^+p \to t\tilde\chi^0_1+X$, $e^-p
\to\bar{t}\tilde\chi^0_1+X$ in case (1) and $e^+p \to
t\tilde\chi^0_1+X$ in case (2), respectively. }
\end{figure}

We plot the LO, NLO QCD corrected cross sections and the
corresponding $K$-factors as the functions of the top squarks mass
$m_{\tilde t_1}$ for the processes $e^+p \to t\tilde\chi^0_1+X$,
$e^-p\to \to\bar{t}\tilde\chi^0_1+X$ in case (1) and $e^+p \to
t\tilde\chi^0_1+X$ in case (2) around the benchmark point as
described in Eq.(\ref{para-susy2}) at the $E_e=50~GeV$ and
$E_p=7~TeV$ LHeC in Fig.\ref{fig7}, respectively. We vary the top squark
mass $m_{\tilde t_1}$ from $500$ to $1000~GeV$ with the other
pMSSM-19 parameters fixed at the benchmark point. We can see that
the NLO QCD corrections generally enhance the cross sections except
for the $e^-p\to e^- \bar{d} \to \bar{t}\tilde\chi^0_1+X$ process in
the region of $570~GeV<m_{\tilde{t}_1}<720~GeV$.

\par
\begin{figure}[htbp]
\includegraphics[scale=0.4]{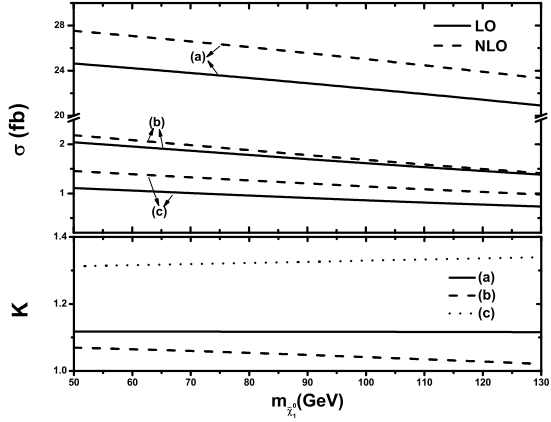}%
\hspace{0in}%
\caption{\label{fig8} The LO, NLO QCD corrected cross sections and
the corresponding $K$-factors versus $m_{\tilde\chi^0_1}$ at the
LHeC with $E_e=50~GeV$ and $E_p=7~TeV$. The curves labeled with
(a), (b), and (c) are for the processes $e^+p \to
t\tilde\chi^0_1+X$, $e^-p \to\bar{t}\tilde\chi^0_1+X$ in case (1)
and $e^+p \to t\tilde\chi^0_1+X$ in case (2), respectively. }
\end{figure}

The LO, NLO QCD corrected cross sections and the corresponding
$K$-factors as the functions of the lightest neutralino mass for the
processes $e^+p \to t\tilde\chi^0_1+X$, $e^-p\to
\to\bar{t}\tilde\chi^0_1+X$ in case (1) and $e^+p \to
t\tilde\chi^0_1+X$ in case (2), at the $E_e=50~GeV$ and $E_p=7~TeV$
LHeC are depicted in Fig.\ref{fig8}, separately. In Fig.\ref{fig8}
we keep all the related pMSSM-19 parameters as the values at the
benchmark point shown in Eq.(\ref{para-susy2}) except the lightest
neutralino mass. The results show that the NLO QCD corrections
always increase the corresponding LO cross sections when
$m_{\tilde{\chi}_1^0}$ varies from $50~GeV$ to $130~GeV$.

\par
\begin{figure}[htbp]
\includegraphics[scale=0.4]{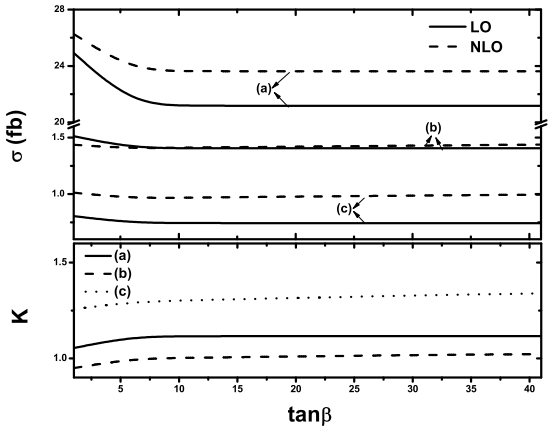}%
\hspace{0in}%
\caption{\label{fig9} The LO, NLO QCD corrected cross sections and
the corresponding $K$-factors versus $\tan\beta$ at the LHeC with
$E_e=50~GeV$ and $E_p=7~TeV$. The curves labeled with (a), (b),
and (c) are for the processes $e^+p \to t\tilde\chi^0_1+X$, $e^-p
\to\bar{t}\tilde\chi^0_1+X$ in case (1) and $e^+p \to
t\tilde\chi^0_1+X$ in case (2), respectively.  }
\end{figure}

We depict the LO, NLO QCD corrected cross sections and the
corresponding $K$-factors versus the ratio of the vacuum expectation
values (VEVs), $\tan\beta$, for the processes $e^+p \to
t\tilde\chi^0_1+X$, $e^-p \to\bar{t}\tilde\chi^0_1+X$ in case (1)
and $e^+p \to t\tilde\chi^0_1+X$ in case (2), at the $E_e=50~GeV$
and $E_p=7~TeV$ LHeC in Fig.\ref{fig9}, respectively. There all the
related pMSSM-19 parameters are fixed at the benchmark point and
remain unchanged except $\tan\beta$. The curves in the figure
demonstrate that the cross sections decrease as $\tan\beta$
increases in the range of $\tan\beta<7$. While when $\tan\beta$ goes
beyond $10$, the cross section is almost independent of $\tan\beta$.

\par
\begin{figure}[htbp]
\includegraphics[scale=0.4]{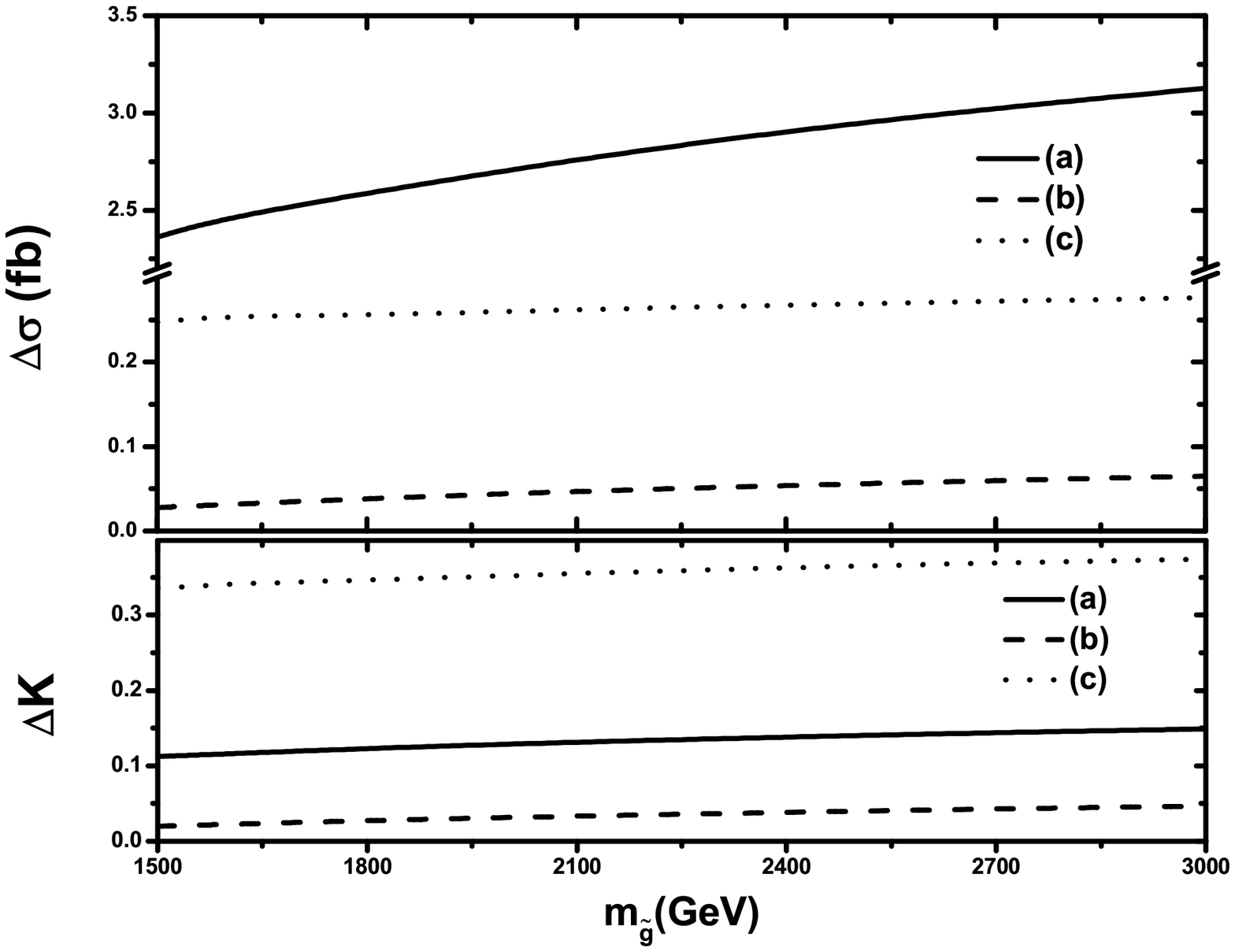}%
\hspace{0in}%
\caption{\label{fig10} The NLO QCD correction ($\Delta\sigma\equiv
\sigma_{NLO}-\sigma_{LO}$) and the corresponding $\Delta K\equiv
\Delta\sigma/\sigma_{LO}$ versus the gluino mass $m_{\tilde g}$ at
the $E_e=50~GeV$ and $E_p=7~TeV$ LHeC. The curves labeled with
(a), (b), and (c) are for the processes $e^+p \to
t\tilde\chi^0_1+X$, $e^-p \to\bar{t}\tilde\chi^0_1+X$ in case (1)
and $e^+p \to t\tilde\chi^0_1+X$ in case (2), respectively. }
\end{figure}

The NLO QCD correction $\Delta\sigma_{NLO}(\equiv
\sigma_{NLO}-\sigma_{LO})$ and the corresponding $\Delta K\equiv
\Delta\sigma_{NLO}/\sigma_{LO}$ versus the gluino mass $m_{\tilde
g}$ at the $E_e=50~GeV$ and $E_p=7~TeV$ LHeC, are demonstrated in
Fig.\ref{fig10} for the processes $e^+p \to t\tilde\chi^0_1+X$,
$e^-p \to\bar{t}\tilde\chi^0_1+X$ in case (1) and $e^+p \to
t\tilde\chi^0_1+X$ in case (2), around the benchmark point with SUSY
parameters in Eq.(\ref{para-susy2}). In these figures
we keep all the related SUSY parameters remain unchanged except
$m_{\tilde g}$. It shows that when the gluino mass runs from $1500~GeV$
to $3500~GeV$, the $\Delta K$ increases slowly for each of the three
curves.

\par
\begin{figure}[htbp]
\includegraphics[width=3.2in,height=2.4in]{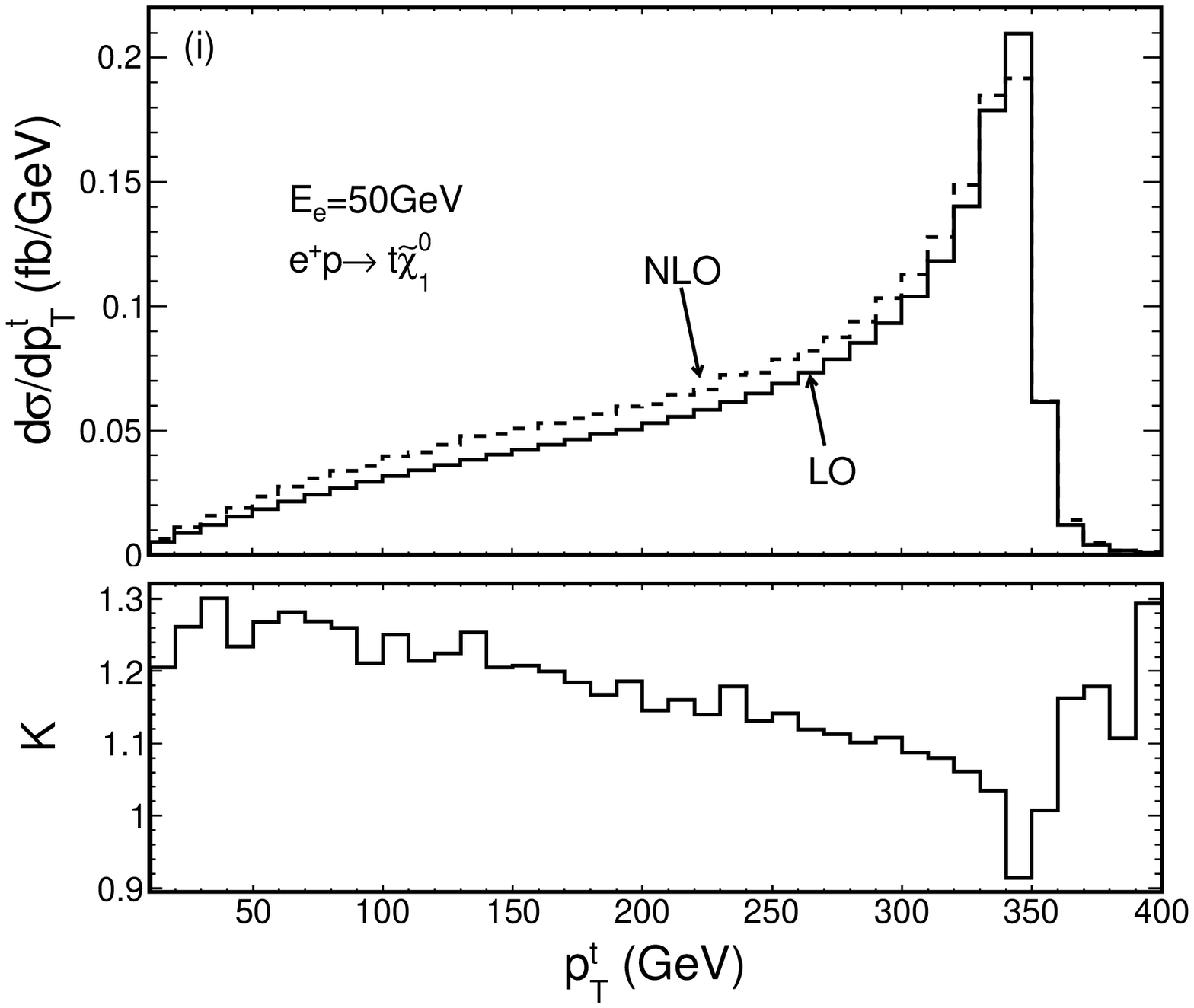}%
\hspace{0in}%
\includegraphics[width=3.2in,height=2.4in]{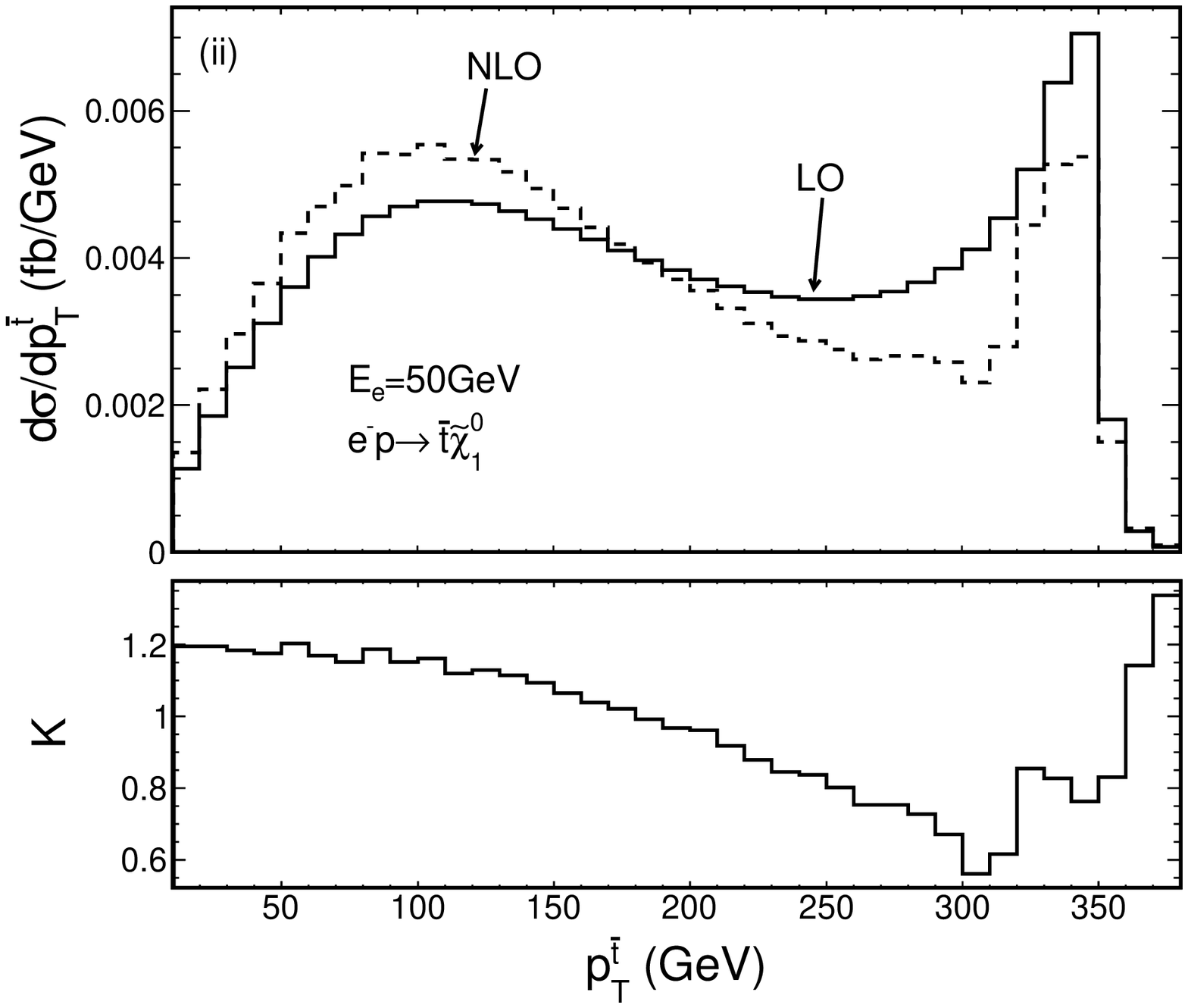}%
\hspace{0in}%
\includegraphics[width=3.2in,height=2.4in]{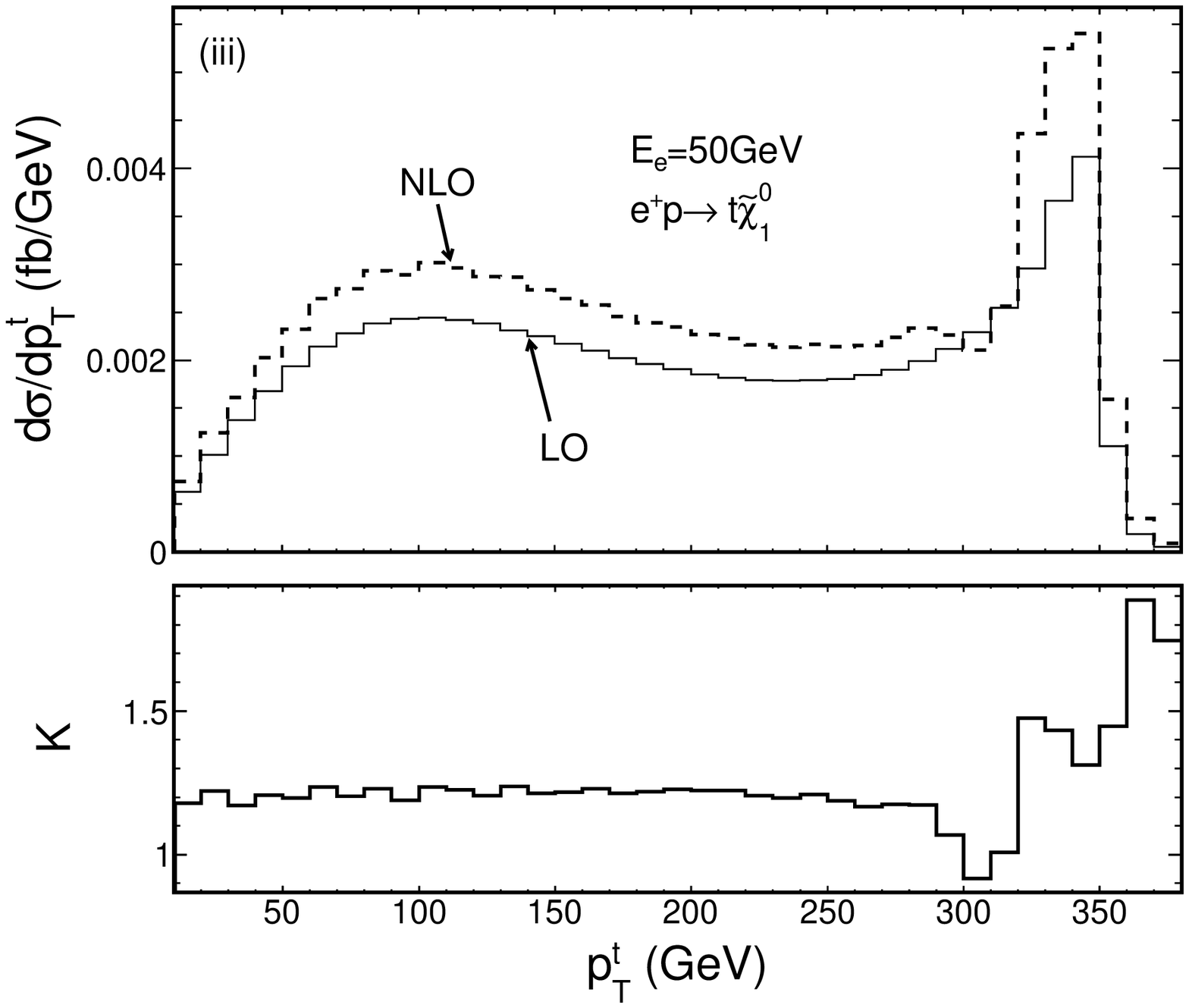}%
\hspace{0in}%
\caption{\label{fig11} The LO, NLO QCD corrected distributions of
the $p_T$ of final (anti-)top quark and the corresponding
$K$-factors $\left(K(p_T)\equiv \frac{d \sigma_{NLO}}{dp_T}/
\frac{d\sigma_{LO}}{dp_T}\right)$ at the benchmark point at the
$E_e=50~GeV$ and $E_p=7~TeV$ LHeC. (i) The process $e^+p \to
t\tilde\chi^0_1+X$ in case (1). (ii) The process $e^-p
\to\bar{t}\tilde\chi^0_1+X$ in case (1). (iii) The process $e^+p
\to t\tilde\chi^0_1+X$ in case (2). }
\end{figure}

We show the transverse momentum distributions of the final
(anti-)top quark and the corresponding $K$-factors for the processes
$e^+p \to t\tilde\chi^0_1+X$, $e^-p \to\bar{t}\tilde\chi^0_1+X$ in
case (1) and $e^+p \to t\tilde\chi^0_1+X$ in case (2), at the
benchmark point mentioned above at the $E_e=50~GeV$ and $E_p=7~TeV$
LHeC in Figs.\ref{fig11}(i), \ref{fig11}(ii), and \ref{fig11}(iii), respectively. The
transverse momentum distributions in Fig.\ref{fig11} show that
there exist peaks located at the position about $p^t_T(p^{\bar t}_T)
\sim 345~GeV$. Those peaks originate from the resonant $\tilde{t}_1$
effects in the $s-$channel diagrams. We can see that the peak on the
NLO transverse momentum distribution curve is lower than that on the
corresponding LO curve in Figs.\ref{fig11}(i) and \ref{fig11}(ii). The
$K$-factor reaches the lowest value at the peak of transverse
momentum distribution, and then leaps to a very high value when
$p_T$ goes up beyond the peak position. Our numerical evaluation
shows that the large values of the $K$-factors in the
$p_T^{t(\bar{t})}>350~GeV$ regions of Fig.\ref{fig11} come from the
large contributions of the real gluon emission processes. We can
conclude that the NLO QCD correction to the transverse momentum
distribution of the $t(\bar{t})$-quark can be very significant in
most of the $p_T$ regions.

\par
In Table \ref{tab-3}, we present some typical numerical results with
single dominance hypotheses of $\lambda_{131}^{\prime} = 0.1$ [case
(1)] and $\lambda_{132}^{\prime} = 0.1$ [case (2)] for two electron
beam energy configurations of $E_e = 50~GeV$ and $E_e = 150~GeV$ for
comparison. With the assumption of integrated luminosities ${\cal
L}_{int} = 20~fb^{-1}$ and $2~fb^{-1}$ for $E_e = 50~GeV$ and $E_e
=150~GeV$, respectively, we can collect about $470$ and $220$ events
for the $e^+ p \to t \tilde{\chi}_1^0 + X$ process in case (1) at the
$E_e = 50~GeV$ and $E_e = 150~GeV$ LHeC, respectively. The
corresponding relative statistic errors are estimated as $4.6\%$ and
$6.7\%$, which are less than the QCD corrections of $11.5\%$
$(K=1.115)$ and $9.7\%$ $(K=1.097)$, respectively. Therefore, the
QCD corrections might be resolvable for a precision measurement of
$e^+ p \to t \tilde{\chi}_1^0 + X$ in case (1). However with the same
luminosity assumptions, the LO and NLO corrected cross sections for
the $e^+p \to t\tilde\chi^0_1+X$ process in case (2) at the $E_e =
50~GeV$ LHeC are less than $1~fb$ and hardly resolvable in high
precision measurement.
\begin{table}
\begin{center}
\begin{tabular}{|c|c|c|c|}
\hline Process  & {$\sigma_{LO}$(fb)} &$\sigma_{NLO}$(fb) & $K$-factor \\
\hline \tabincell{c}{$e^+p \to t\tilde\chi^0_1+ X$\\in Case (1)}
& 101.91(1) $\parallel$ 21.240(3) & 111.8(1) $\parallel$ 23.69(2) & 1.097 $\parallel$ 1.115 \\
\hline \tabincell{c}{$e^-p\to \bar t\tilde\chi^0_1+ X $\\in Case
(1)}
& 27.621(5) $\parallel$ 1.4101(3) & 28.65(5) $\parallel$ 1.441(2) & 1.037 $\parallel$ 1.022 \\
\hline \tabincell{c}{$e^+p \to t\tilde\chi^0_1 + X$ \\ in Case (2)}
& 13.820(2) $\parallel$ 0.7447(1) & 14.85(2) $\parallel$ 0.998(1) & 1.075 $\parallel$ 1.340 \\
\hline
\end{tabular}
\end{center}
\caption{ \label{tab-3} The numerical results of the LO and NLO QCD
corrected total cross sections for the processes $e^+p \to
t\tilde\chi^0_1+X$, $e^-p \to\bar{t}\tilde\chi^0_1+X$ in case (1)
and $e^+p \to t\tilde\chi^0_1+X$ in case (2) at the benchmark point
at the $E_e=50~GeV$ and $E_e=150~GeV$ LHeC. The data on the left
(right) side of the notation $\parallel$ correspond to the results
at $E_e=150~GeV$ ($E_e=50~GeV$) LHeC. }
\end{table}

\vskip 5mm
\section{Summary}
\par
In this paper, we calculate the NLO QCD corrections to the single
(anti-)top quark production associated with a lightest neutralino in the
MSSM with $R$-parity breaking coupling at the LHeC. In our
calculations, we include completely all possible resonant and
nonresonant diagrams and do not assume the production and decay
factorization for the top squarks. The UV divergences are eliminated
in the complex top squark mass scheme and the IR divergences are
canceled by adopting both the TCPSS and DPS methods to make cross-check. 
The numerical calculations are carried out with the chosen
SUSY parameters in the pMSSM-19, which is compatible with current
experimental bounds. We investigate the dependence of the LO and the
NLO QCD integrated cross sections on the top squark, neutralino, and
gluino masses around the benchmark point. We also present the LO and
the NLO QCD corrected distributions of the transverse momenta of
final (anti-)top quark; they show that the impacts of the NLO QCD
corrections might be resolvable in high precision measurement.

\vskip 5mm
\par
\noindent{\large\bf Acknowledgments:} This work was supported in
part by the National Natural Science Foundation of China
(Grants No. 11075150, No. 11005101, No. 11275190), and the Fundamental Research
Funds for the Central Universities (Grant No. WK2030040024).


\vskip 10mm
\begin{thebibliography}{99}
\bibitem{exbounds1}
ATLAS Collaboration, Report No. ATLAS-CONF-2013-007;
ATLAS Collaboration, Report No. ATLAS-CONF-2013-024;
ATLAS Collaboration, Phys. Lett. B {\bf 720} 13 (2013);
ATLAS Collaboration, J. High Energy Phys. 02 (2013) 0951.
\bibitem{exbounds2}
CMS Collaboration, Report No. CMS PAS SUS-12-024;
CMS Collaboration, Report No. CMS PAS SUS-13-007;
CMS Collaboration, Phys. Rev. D {\bf 87} 052006 (2013).
\bibitem{SusyStatus}
S. Kraml, arXiv:hep-ph/1206.6618.
\bibitem{naturesusy}
Michele Papucci, Joshua T. Ruderman, and Andreas Weiler, Report No. DESY-11-193,
CERN-PH-TH/265, J. High Energy Phys. 09 (2012) 035.
\bibitem{higgsfit}
Junjie Cao, Zhaoxia Heng, Jin Min Yang, and Jingya Zhu,
J. High Energy Phys. 10 (2012) 079; F. Mahmoudi, A. Arbey, M. Battaglia, and A.
Djouadi, arXiv:hep-ph/1211.2794; Lorenzo Basso and  Florian Staub,
Phys. Rev. D {\bf 87} 015011 (2013); Vernon Barger, Peisi Huang, Muneyuki Ishida, and 
Wai-Yee Keung, Phys. Rev. D {\bf 87} 015003 (2013); Jiwei Ke, Hui Luo, Ming-Xing
Luo {\it et al.}, Phys. Lett. B {\bf 723} 113 (2013); Zhaoxia Heng,
arXiv:hep-ph/1210.3751; Marcela Carena, Stefania Gori, Ian Low {\it
et al.}, J. High Energy Phys. 02 (2013) 114.
\bibitem{pMSSM1}
Matthew W. Cahill-Rowley, {\it et al.}, Eur. Phys. J. C {\bf 72} 2156 (2012).
\bibitem{Martin}
S. P. Martin, Phys. Rev. {\bf D 54}, 2340 (1996);
P. F. P\'{e}rez and S. Spinner, Phys. Lett. B {\bf 673}, 251 (2009).
\bibitem{rp1}
P. Fayte, Phys. Lett. B {\bf 69}, 489 (1977).
\bibitem{rp2}
G. R. Farrar, P. Fayte, Phys. Lett. B {\bf 76}, 575 (1978).
\bibitem{superpotential_1}
S. Weinberg, Phys. Rev. D {\bf 26}, 287 (1982).
\bibitem{superpotential_2}
N. Sakai and  T. Yanagida, Nucl. Phys. {\bf B197}, 533 (1982).
\bibitem{superpotential_3}
R. Barbier, C. Berat, M. Besancon {\it et al.}, Phys. Rep. {\bf 420}, 1 (2005).
\bibitem{LBviolation}
L.E. Ibanez and  G.G. Ross, Nucl. Phys. {\bf B368}, 3 (1992).
\bibitem{LHeC}
J. L. Abelleira Fernandez {\it et al.}, J. Phys. G {\bf 39} 075001;
J. L. Abelleira Fernandez, C. Adolphsen {\it et al.}, arXiv:hep-ex/1211.5102;
LHeC web page, http://www.lhec.org.uk.
\bibitem{hera1}
T. Kon, T. Kobayashi, and S. Kitamura, Phys. Lett. B {\bf 333}, 263 (1994);
T. Kobayashi, S. Kitamura, and T. Kon, Int. J. Mod. Phys. A {\bf 11}, 1875 (1996);
T. Kon, T. Kobayashi, and S. Kitamura, Phys. Lett. B {\bf 376}, 227 (1996);
T. Kon, T. Kobayashi, Phys. Lett. B {\bf 409}, 265 (1997).
\bibitem{hera2}
H. Dreiner and  P. Morawitz, Nucl. Phys. {\bf B503}, 55 (1997).
\bibitem{hera3}
Jihn E. Kim and  P. Ko, Phys. Rev. D {\bf 57}, 489 (1998).
\bibitem{hera4}
J. Ellis, S. Lola, and K. Sridhar, Phys. Lett. B {\bf 408}, 252 (1997).
\bibitem{hera5}
Z. Kunszt, and W. J. Stirling, Z. Phys. C {\bf 75}, 453 (1997).
\bibitem{work1}
Wei Hong-Tang, Zhang Ren-You, Guo Lei, Han Liang, Ma Wen-Gan {\it et al.}, 
J. High Energy Phys. 07 (2011) 003.
\bibitem{work2}
M. Arai, K. Huitu, S.K. Rai, and K. Rao, J. High Energy Phys. 08 (2010) 082.
\bibitem{work3}
M.A. Bernhardt, H.K. Dreiner, S. Grab, and P. Richardson, Phys. Rev. D {\bf 78}, 015016 (2008).
\bibitem{work4}
M. Chemtob and  G. Moreau, Phys. Rev. D {\bf 61}, 116004 (2000).
\bibitem{work5}
Siba Prasad Das, Amitava Datta, and Monoranjan Guchait, Phys. Rev. D {\bf 70}, 015009 (2004).
\bibitem{rpvbounds}
R. Barbier {\it et al.}, Phys. Rep. {\bf 420}, 1 (2005).
\bibitem{bounds}
F. Borzumati, and J.S. Lee, Phys. Rev. D {\bf 66}, 115012 (2002).
\bibitem{cms}
A. Denner, S. Dittmaier, M. Roth, and L.H. Wieders, Nucl. Phys. {\bf B724}, 247 (2005);
A. Denner, S. Dittmaier, M. Roth, and D. Wackeroth, Nucl. Phys. {\bf B560}, 33 (1999).
\bibitem{shiftgs}
  W. Beenakker, R. H\"opker, and P. M. Zerwas, Phys. Lett. {\bf B378}, 159 (1996);
  W. Beenakker, R. H\"opker, T. Plehn, and P.M. Zerwas, Z. Phys. C {\bf 75}, 349 (1997).
\bibitem{os}
A. Denner,  Fortschr. Phys. {\bf 41}, 307 (1993).
\bibitem{msbar}
W.J. Marciano, Phys. Rev. D {\bf 29}, 580 (1984).
\bibitem{sol4pi}
A. Denner and S. Dittmaier, Nucl. Phys. {\bf B844}, 199 (2011).
\bibitem{irsafe}
G. t'Hooft and M. Veltman, Nucl. Phys. {\bf B153}, 365 (1979);
A. Denner, U. Nierste, and R. Scharf, Nucl. Phys. {\bf B367}, 637 (1991);
A. Denner and S. Dittmaier, Nucl. Phys. {\bf B658}, 175 (2003).
\bibitem{dpf}
P.F. Duan, R.Y. Zhang, W.G. Ma, L. Guo, and Y. Zhang, J. Phys. G {\bf 39}, 105002 (2012).
\bibitem{Decouple}
H. K. Dreiner, S. Grab, M. Kr\"amer, and M. K. Trenkel, Phys. Rev. D {\bf 75}, 035003 (2007).
\bibitem{KLN_1}
T. Kinoshita,  J. Math. Phys. {\bf 3}, 650 (1962).
\bibitem{KLN_2}
T.D. Lee and  M. Nauenberg, Phys. Rev. {\bf 133}, B1549 (1964).
\bibitem{TCPSS}
B.W. Harris and  J.F. Owens, Phys. Rev. D {\bf 65}, 094032 (2002).
\bibitem{dipole}
Stefano Catani and  Michael H. Seymour, Nucl. Phys. {\bf B485}, 291 (1997).
\bibitem{cteq_1}
J. Pumplin {\it et al.}, J. High Energy Phys. 07 (2002) 012.
\bibitem{cteq_2}
D. Stump {\it et al.}, J. High Energy Phys. 10 (2003) 046.
\bibitem{Amsler}
C. Amsler {\it et al.}, Phys. Lett. B {\bf 667}, 1 (2008);
K. Nakamura {\it et al.}, J. Phys. G {\bf 37}, 075021 (2010).
\bibitem{pdglive}
J. Beringer {\it et al.}, Phys. Rev. D {\bf 86}, 010001 (2012);
http://pdglive.lbl.gov/listings1.brl?quickin=Y.1.
\bibitem{sigmis1}
For recent papers, see, e.g. Manimala Chakraborty, Utpal
Chattopadhyay, and Rohini M. Godbole, Phys. Rev. D {\bf 87}, 035022
(2013); Howard Baer {\it et al}, Phys. Rev. D {\bf 87}, 035017 (2013);
Gautam Bhattacharyya and Tirtha Sankar Ray, Phys. Rev. D {\bf 87},
015017 (2013); Motoi Endo {\it et al}, Phys. Rev. D {\bf 85}, 095006
(2012); E. Gabrielli, K. Kannike, B. Mele, A. Racioppi, and M.
Raidal, Phys. Rev. D {\bf 86}, 055014 (2012); Pran Nath,
Int. J. Mod. Phys. {\bf A27} 1230029 (2013).
\bibitem{conshiggs}
Ernesto Arganda, J. Lorenzo Diaz-Cruz, Alejandro Szynkman, and
Eur. Phys. J. C {\bf 73} 2384 (2013); Wolfgang Altmannshofer, Marcela Carena {\it
et al.}, J. High Energy Phys. 01 (2013) 160; Stephen F. King, Margarete
Muhlleitner {\it et al.}, Nucl. Phys. {\bf B870} (2013).
\bibitem{consDM}
For recent papers, see, e.g., F. Mahmoudi, A. Arbey, and M. Battaglia,
arXiv:hep-ph/1211.2795; Alex Drlica-Wagner,
arXiv:astro-ph.HE/1210.5558; D. P. Roy, arXiv:hep-ph/1211.3510;
Leszek Roszkowski, Enrico Maria Sessolo, and Yue-Lin Sming Tsai,
Phys.Rev. D {\bf 86} 095005 (2012); C. Strege, et al,
{\it et al.} J. Cosmol. Astropart. Phys. 03 (2012) 030.
\bibitem{consfo}
For recent papers, see, e.g., F. Mahmoudi, and T. Hurth.,
arXiv:hep-ph/1211.2796; A. Arbey {\it et al}, Phys. Rev. D {\bf 87},
035026 (2013); Jonathan L. Feng, Konstantin T. Matchev, and David
Sanford, Phys. Rev. D {\bf 85}, 075007 (2012); A. Arbey and M.
Battaglia,  Eur. Phys. J. C {\bf 72} 1906 (2012); M.W.
Cahill-Rowley, J. L. Hewett, A. Ismail, and T. G. Rizzo,
arXiv:1211.1981.
\bibitem{CMS-A}
CMS Collaboration, Report No. CMS PAS Hig-13-005.
\bibitem{Atlas-A}
ATLAS Collaboration, Phys. Lett. B {\bf 716} 1 (2012).
\bibitem{Atlas-B}
ATLAS Collaboration, Report No. ATLAS-CONF-2013-034.
\bibitem{Atlas-C}
ATLAS Collaboration, arXiv:1211.6956v2.
\bibitem{atlas}
https://twiki.cern.ch/twiki/bin/view/AtlasPublic/CombinedSummaryPlots$\#$SusyDirect  \\
StopSummary.
\bibitem{Feng}
https://twiki.cern.ch/twiki/bin/view/CMSPublic/PhysicsResultsSUS.
\bibitem{softsusy}
B. C. Allanach, Comput. Phys. Commun. {\bf 143}, 305 (2002).
\bibitem{susyhit}
A. Djouadi, M.M. Muhlleitner, and M. Spira, Acta Phys. Pol. B {\bf 38}, 635 (2007).
\bibitem{maddipole}
R. Frederix, T. Gehrmann, and N. Greiner, J. High Energy Phys. 06 (2010) 086;
R. Frederix, T. Gehrmann, and N. Greiner, J. High Energy Phys. 09 (2008) 122.
\bibitem{maddipole2}
S. Catani, S. Dittmaier, M. H. Seymour, and Z. Trocsanyi, Nucl. Phys. {\bf B627}, 189 (2002).

\end{thebibliography}
\end{document}